\documentclass[fleqn,usenatbib]{mnras}

\usepackage[T1]{fontenc}
\usepackage{ae,aecompl}

\usepackage{graphicx}	
\usepackage{amsmath}	
\usepackage{color}

\usepackage{mathptmx}
\usepackage{txfonts}

\newcommand{\hmpc}{h^{-1}{\rm Mpc}}

\newcommand{\kms}{\;{\rm km}\,{\rm s}^{-1}}

\newcommand{\msolar}{\;{\rm M}_{\odot}}

\newcommand{\gizmo}{{\sc Gizmo}}
\newcommand{\swift}{{\sc Swift}}
\newcommand{\mufasa}{{\sc Mufasa}}
\newcommand{\simba}{{\sc Simba}}
\newcommand{\velociraptor}{{\sc VELOCIraptor}}

\newcommand{\nojet}{{\tt NoJet}}

\title{Cosmological baryon transfer in the \simba{} simulations}
\author[Borrow et al.]{
Josh Borrow$^{1}$,
Daniel Angl\'es-Alc\'azar$^{2, 3}$ \&
Romeel Dav\'e$^{2, 4, 5, 6}$
\\
\\$^1$ Institute for Computational Cosmology, Department of Physics, University of Durham, South Road, Durham, DH1 3LE, UK
\\$^2$ Center for Computational Astrophysics, Flatiron Institute, 162 Fifth Avenue, New York, NY 10010, USA 
\\$^3$ Department of Physics, University of Connecticut, 196 Auditorium Road, U-3046, Storrs, CT 06269-3046, USA
\\$^4$ Institute for Astronomy, University of Edinburgh, Royal Observatory, Edinburgh EH9 3HJ, UK
\\$^5$ University of the Western Cape, Bellville, Cape Town 7535, South Africa
\\$^6$ South African Astronomical Observatories, Observatory, Cape Town 7925, South Africa
}

\begin{document}

\maketitle

\begin{abstract}We present a framework for characterizing the large scale movement of baryons
relative to dark matter in cosmological simulations, requiring only the
initial conditions and final state of the simulation. This is performed using
the {\it spread metric} which quantifies the distance in the final conditions
between initially neighbouring particles, and by analysing the baryonic
content of final haloes relative to that of the initial Lagrangian regions
defined by their dark matter component. Applying this framework to the
\simba{} cosmological simulations, we show that 40\% (10\%) of cosmological
baryons have moved $> 1\hmpc{}$ ($3\hmpc{}$) by $z=0$, due primarily to
entrainment of gas by jets powered by AGN, with baryons moving up to
$12\hmpc{}$ away in extreme cases. Baryons
decouple from the dynamics of the dark matter component due to hydrodynamic
forces, radiative cooling, and feedback processes. As a result, only 60\% of
the gas content in a given halo at $z=0$ originates from its Lagrangian
region, roughly independent of halo mass. A typical halo in the mass range
$M_{\rm vir} = 10^{12}$--$10^{13}\msolar$ only retains 20\% of the gas
originally contained in its Lagrangian region. We show that up to 20\% of the
gas content in a typical Milky Way mass halo may originate in the region
defined by the dark matter of another halo. This {\it inter-Lagrangian baryon
transfer} may have important implications for the origin of gas and metals in
the circumgalactic medium of galaxies, as well as for semi-analytic models of
galaxy formation and “zoom-in" simulations.\end{abstract}


\section{Introduction}
\label{sec:introduction}

Cosmological simulations are an important tool to study the evolution of the
universe. Mass elements of various matter components are tracked over cosmic
time under the influence of gravity and other forces until a desired
redshift, where the distribution of matter can be compared to observations.
The earliest simulations included only dark matter acting under gravity
\citep[see e.g.][]{Frenk1988, Springel2005a}, which remains an important
approach to this day because such simulations are computationally efficient
and can model very large volumes required for, e.g., dark energy studies
\citep{Knabenhans2019}. However, such simulations do not directly model the
observable component. As such, techniques such as semi-analytic models (SAMs)
have been developed \citep{FrenkWhite1990,Kauffmann1996,Somerville1998} to
populate dark matter haloes with galaxies \citep[see e.g.][for modern examples
of SAM frameworks]{Porter2014, Henriques2015, Somerville2015b, Lacey2016}.
Crucially, it has been recognized that feedback processes from the formation
of stars and black holes have an important effect on the resulting observable
baryonic component, though they have a small effect on the collisionless dark
matter. Such feedback often takes the form of large-scale winds that eject
substantial amounts of gas from galaxies due to energetic input from young
stars, supernovae, and active galactic nuclei (AGN). This gas can then be
deposited far out in the intergalactic medium (IGM), remain as halo gas in
the Circumgalactic Medium (CGM), or be re-accreted in `wind recycling'
\citep{Oppenheimer2010, Christensen2016, AnglesAlcazar2017, Hafen2019,
Christensen2018}. This cycling of baryons is an integral part of modern
galaxy formation theory, and is believed to be a key factor in establishing
the observed properties of both galaxies and intergalactic gas
\citep{Somerville2015}.

With advancing computational speed and algorithmic developments, it has
become possible to run full hydrodynamical models of the universe that
explicitly track the baryonic component \citep[e.g.][]{Hernquist1989,
Teyssier2002, Springel2005b}. Beyond modelling hydrodynamical processes,
sub-grid prescriptions have been implemented in order to cool the gas and
produce stars, with increasing levels of refinement and sophistication
\citep[e.g.][]{Revaz2012, Vogelsberger2014, Schaye2015, Hopkins2018}. Using
these models it is now possible to reproduce many of the key observed
properties of galaxies at a range of cosmic epochs. Modern galaxy formation
simulations typically include radiative cooling, chemical enrichment, star
formation, stellar feedback, and AGN feedback. Despite playing a critical
role in regulating galaxy growth \citep{Naab2017}, feedback remains poorly
understood. These models must prevent too much star formation, as well as the
`overcooling problem', suffered by the earliest hydrodynamical simulations
\citep{Dave2001,Balogh2001}.

Feedback processes also transport baryons far from their originating dark
matter haloes. Early observational evidence for this was that the diffuse
intergalactic medium at high redshift is enriched with metals produced by
supernovae, requiring winds with speeds of hundreds of $\kms{}$ to be ejected
ubiquitously \citep[e.g.][]{Aguirre2001, Springel2003, Oppenheimer2006}. More
recently, feedback from AGN is seen to eject ionised and molecular gas
outflows with velocities exceeding 1000 $\kms{}$
\citep[e.g.][]{Sturm2001, Greene2012, Maiolino2012, Zakamska2016}. It has
long been known that some AGN also power jets, carrying material out at
relativistic velocities \citep{Fabian2012}. These processes decouple the
baryonic matter from the dark matter on cosmological scales, which could
potentially complicate approaches to populating dark matter simulations with
baryons. Hence it is important to quantify the amount of baryons that are
participating in such large-scale motions, within the context of modern
galaxy formation models that broadly reproduce the observed galaxy
population.

This paper thus examines the large-scale redistribution of baryons relative to the
dark matter, using the \simba{} cosmological simulations that include kinetic
feedback processes which plausibly reproduces the observed galaxy population
\citep{Dave2019}. To do this, we pioneer a suite of tools to compare the
initial and final location of baryons relative to their initial `Lagrangian
region', defined as the region in the initial conditions that collapses into
a given dark matter halo. In classical galaxy formation theory, the baryons
follow the dark matter into the halo, and only then significantly decouple
thanks to radiative processes; this would result in the baryons lying mostly
within the Lagrangian region of the halo. However, outflows can disrupt this
process, and result in the transfer of baryons outside the Lagrangian region
or even transfer \emph{between} Lagrangian regions. It is these effects we
seek to quantify in this work.

The importance of ejecting baryons and the resulting transfer of material to
other galaxies was highlighted using recent cosmological `zoom-in'
simulations from the FIRE project \citep{Hopkins2014,Hopkins2018}. Tracking
individual gas resolution elements in the simulations,
\citet{AnglesAlcazar2017} showed that gas ejected in winds from one galaxy
(often a satellite) can accrete onto another galaxy (often the central) and
fuel in-situ star formation. This mechanism, dubbed `intergalactic transfer',
was found to be a significant contributor to galaxy growth. The galaxies that
provided intergalactic transfer material often ended up merging with the
central galaxy by $z=0$, with their mass contribution via winds greatly
exceeding that of the merger events. However, this work did not examine the
extent to which galactic winds can push gas to larger scales and connect
individual haloes at $z=0$, since it is not feasible to examine this in
zoom-in simulations that by construction focus on modelling a single halo.

In this work, we consider matter flows in a large cosmological volume ($50
\hmpc{}$) using the \simba{} simulations \citep{Dave2019}, whose star
formation feedback employs scalings from FIRE, and whose black hole model
includes various forms of AGN feedback including high-velocity jets. More
generally, we present a framework for analysing the relative motion of dark
matter and baryons on large scales due to hydrodynamic and feedback
processes. With this, we quantify the large scale gas flows out of Lagrangian
regions into the surrounding IGM and the importance of `inter-Lagrangian
transfer' in galaxy evolution.

The remainder of this paper is organised as follows: in \S\ref{sec:simba}, we
discuss the \simba{} simulation suite that is used for analysis; in
\S\ref{sec:feedbackmetrics}, we discuss a distance-based metric for the
investigation of feedback strength; in \S\ref{sec:transfer}, we discuss
halo-level metrics based on Lagrangian regions to study inter-Lagrangian
transfer; in \S\ref{sec:convergence} we discuss the convergence of the
method; and in \S\ref{sec:conclusions} we conclude and summarise the results.
\section{The \simba{} Simulation Suite}
\label{sec:simba}

\subsection{Code and sub-grid model}

This work uses the \simba{} simulation suite \citep{Dave2019}, which inherits
a large amount of physics from \mufasa{} \citep{Dave2016}. \simba{} uses a
variant of the GIZMO code \citep{Hopkins2015}, with the Meshless-Finite-Mass
(MFM) hydrodynamics solver using a cubic spline kernel with
64 neighbours. The gravitational forces are solved using the Tree-PM method
as described in \citet{Springel2005b} for Gadget-2, of which GIZMO is a
descendent. In the $50 \hmpc{}$, $512^3$ particle box used here, the mass
resolution for the gas elements is $1.7\times10^7h^{-1}$ M$_\odot$, and for
the dark matter is $7\times10^7h^{-1}$ M$_\odot$. The cosmology used in
\simba{} is consistent with results from \citet{PlanckCollaboration2016},
with $\Omega_\Lambda = 0.7$, $\Omega_{\rm m} = 0.3$, $\Omega_{\rm b} =
0.048$, $H_0 = 68$ km s$^{-1}$, $\sigma_8=0.82$, and $n_s=0.97$.

On top of this base code, the \simba{} sub-grid model is implemented. This
model is fully described in \citet{Dave2019}, but it is summarised here.
Radiative cooling and photoionisation are included from Grackle-3.1
\citep{Smith2016}. Stellar feedback is modelled using decoupled two-phase
winds that have 30\% of their ejected particles set at a temperature given by
the supernova energy minus the kinetic energy of the wind. The mass loading
factor of these winds scales with stellar mass using scalings from
\citet{AnglesAlcazar2017}, obtained from particle tracking in the FIRE
zoom-in simulations.

Black hole growth is included in \simba{} using the torque-limited accretion
model from \citet{AnglesAlcazar2017b} for cold gas and \citet{Bondi1952}
accretion for the hot gas. The AGN feedback model includes both kinetic winds
and X-ray feedback. At high Eddington ratios ($f_{\rm Edd} > 0.02$) or low
black holes mass ($M_{\rm BH} < 10^{7.5}$ M$_\odot$), the radiative-mode
winds are high mass-loaded and ejected at interstellar medium (ISM)
temperature with velocities $\lesssim 10^3 \kms{}$. At low Eddington ratios
and high black hole mass, the jet-mode winds are ejected at velocities
approaching $\sim 10^4\kms{}$. We refer the interested reader to the full
description of this feedback model in \citet{Dave2019}.

In addition to the fiducial model, we also use two comparison models. The
first, described as \nojet{}, includes all of the \simba{} physics but has
the high-energy black hole jet-mode winds disabled. All other star formation
and AGN feedback is included. The second, described as non-radiative, uses
the same initial conditions as the fiducial model but only includes
gravitational dynamics and hydrodynamics, i.e. without sub-grid models. This
latter simulation was performed with the {\sc Swift} simulation code
\citep{Schaller2016} using a Density-Entropy Smoothed Particle Hydrodynamics
(SPH) solver as it performs orders of magnitude faster than the original
GIZMO code \citep{Borrow2018}. The use of this hydrodynamics model, over the
MFM solver, will have a negligible effect on the quantities of interest in
this paper, as it has been shown that such a solver produces haloes of the
same baryonic mass when ran in non-radiative mode \citep[see
e.g.][]{Sembolini2016}.

\subsection{Defining haloes}

Haloes are defined using a modified version of the Amiga Halo Finder
\citep[AHF, ][]{Gill2004, Knollmann2009} presented in \citet{Muratov2015}.
This spherical overdensity finder determines the halo centers by using a
nested grid, and then fits parameters based on the Navarro-Frenk-White
\citep[NFW, ][]{Navarro1995} profile. Here we define the virial radius,
$R_{\rm vir}$, as the spherical overdensity radius retrieved from AHF consistent
with \citet{Bryan1998}. Substructure search was turned off, such that the
code only returned main haloes.

\subsection{Defining Lagrangian regions}

The Lagrangian region (LR) associated with a halo is the volume in the initial conditions
that contains the dark matter that will eventually collapse to form that halo.

Many methods exist for defining Lagrangian regions \citep[see e.g. ][ for a
collection of methods]{Onorbe2014}. In this work the Lagrangian regions are
defined in the following way:
\begin{enumerate}
	\item Find all haloes at redshift $z=0$, and assign them a unique halo ID.

    \item For each halo, match the particles contained within it with those
		  in the initial conditions. These particles are then assigned a Lagrangian
		  region ID that is the same as this halo ID, with particles outside of haloes
		  (and hence Lagrangian regions) assigned an ID of -1. This defines the initial
		  Lagrangian regions based on the dark matter.

	\item In some cases, discussed below, fill in the holes in this Lagrangian
		  region by using a nearest-neighbour search. In the fiducial case, skip
		  this step (see \S \ref{sec:convergence}).

	\item For every gas particle in the initial conditions, find the nearest dark
	      matter neighbour. This gas particle is assigned to the same Lagrangian
	      region as that dark matter particle.
\end{enumerate}
In this way, Lagrangian regions contain all dark matter particles that end up
within $R_{\rm vir}$ of each halo at $z=0$, by definition, as well as the
baryons that should also in principle collapse into the corresponding halo.
In \S \ref{sec:convergence}, we explore alternative definitions of LRs and
their impact in our results.
\section{Quantifying Baryon Redistribution}
\label{sec:feedbackmetrics}

Feedback is a complex process that impacts a wide range of baryonic
observables, from the galaxy stellar mass function, to galaxy sizes, to the
density profiles of galaxies \citep[e.g.][]{Angles-Alcazar2014, Nelson2015,
Hellwing2016, BenitezLlambay2018}. It is interesting, therefore, to develop
tools to study the global effects of feedback directly, as a complement to
the many indirect constraints obtainable from comparing to astrophysical
observables. Here we describe the {\it spread metric} as a general tool to
examine the redistribution of baryons via feedback relative to the underlying
dark matter distribution.

\subsection{The Spread Metric}

\begin{figure}
    \centering
    \includegraphics[width=\columnwidth]{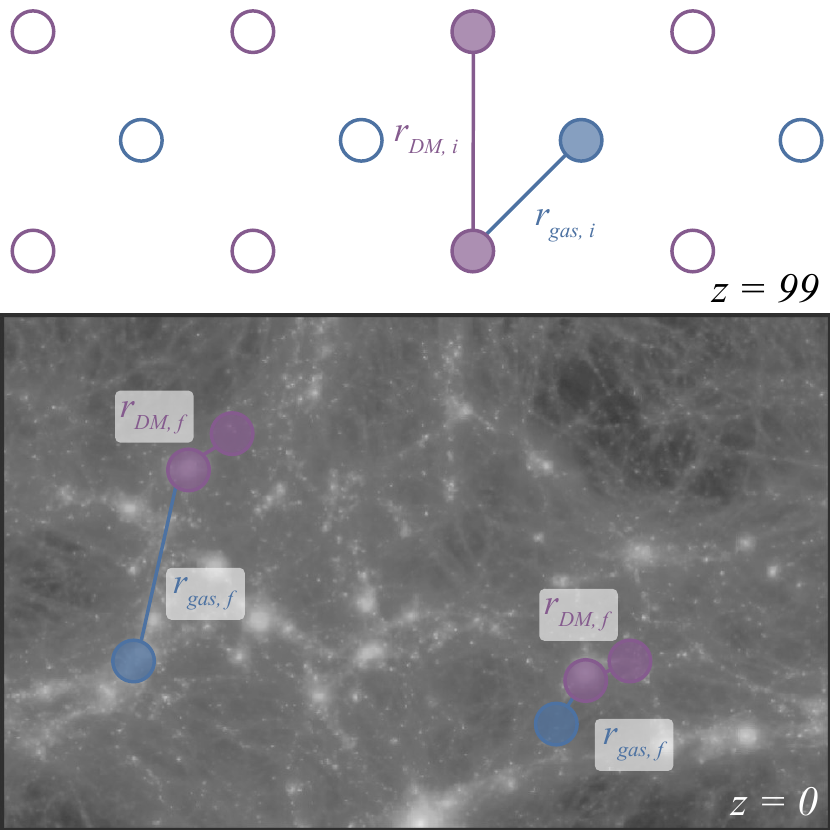}
    \vspace{-0.5cm}
    \caption{Illustration of the matching procedure between initial and final
    conditions to define the spread metric. Gas particles are shown in
    blue, with dark matter particles shown in purple. The top panel shows the
    $z=99$ initial conditions, where every particle finds its nearest dark
    matter neighbour. The bottom panel shows the distances between those
    particles at $z=0$. For our fiducial results, each particle is matched to
    the three nearest neighbours at $z=99$ and the spread metric is computed
    as the median of the corresponding distances at $z=0$ (see text for
    details).}
    \label{fig:kspafigsmall}
\end{figure}

Our approach to quantifying the large-scale impact of feedback is to develop
a simple and robust metric that directly captures the displacement of gas
due to feedback. This {\it spread metric}, illustrated in Fig.
\ref{fig:kspafigsmall}, works as follows:

\begin{enumerate} 
	\item For every gas particle $i$ in the initial conditions, find the nearest
          $n$ dark matter neighbours $j$ (with $n=3$ for our fiducial results).
	\item In the final conditions at $z=0$, match all remaining baryonic particles
	      with their initial conditions progenitor (in this case, stars are
	      matched with their gas particle progenitor).
    \item Find the distance $r_{ij}$ between particles $i$ and $j$ in the
          final conditions.
    \item The spread metric for particle $i$, denoted $S_{i}$, is given by the \emph{median}
          of the $n$ original dark matter neighbour distances $r_{ij}$.
\end{enumerate}

\begin{figure}
    \centering
    \includegraphics{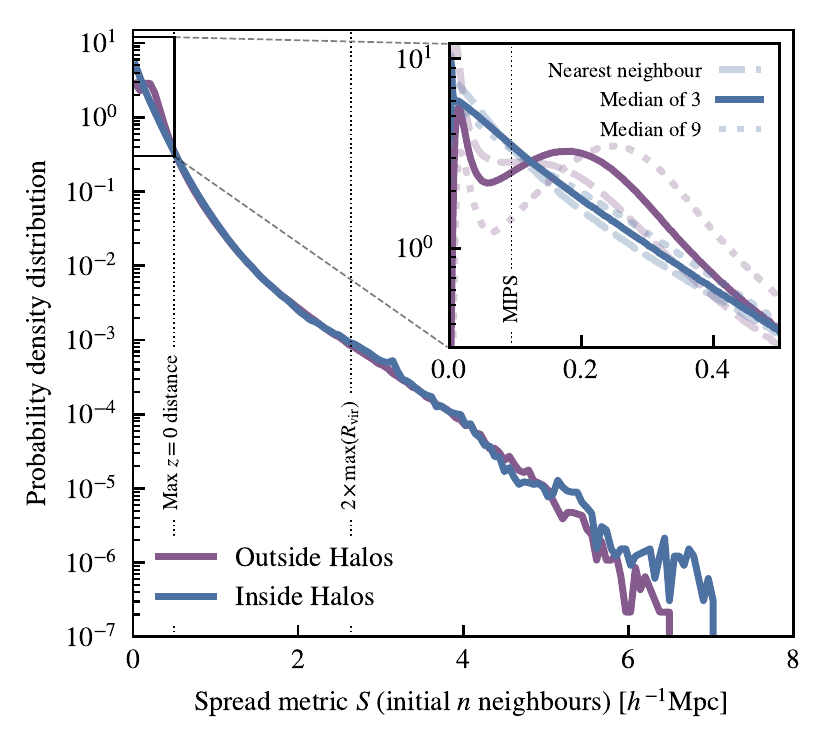}
    \vspace{-0.7cm}
    \caption{The redshift $z=0$ spread metric distribution for the dark
    matter component in the full \simba{} model. The distribution is split
    between particles that lie within haloes (blue) and outside haloes
    (purple), with this being an approximately even split at $z=0$. Vertical
    dotted lines indicate the maximal distance between any two nearest dark
    matter particles at $z=0$ ($\sim 0.5\hmpc{}$) and twice the maximal
    virial radius of any halo in the box (${\rm max}(R_{\rm vir}) \sim
    1.3\hmpc{}$). The inset figure shows the inner $0.5\hmpc{}$ of the
    distribution, with the mean inter-particle separation in the initial
    conditions (MIPS $\sim 0.1\hmpc$) indicated by the vertical dotted line.
    The fainter lines show how the spread metric changes when taking the
    median over a different number of initial nearest neighbours.
    This figure shows that initially neighbouring dark matter particles
    can be spread out to $7\hmpc{}$ due to gravitational dynamics alone.}
    \label{fig:dmonlyspread}
\end{figure}

The spread metric is introduced to measure the net displacement of baryons over
cosmic time. This is somewhat difficult to do in practice, as to measure the
net movement of particles we require a reference point. We take that
reference point to be the initially neighbouring dark matter particle as to
respect the Lagrangian nature of the simulation. This is different to taking the
relative motion of the particle compared to its initial point in co-moving space
as it ensures that there is zero `spread' in bulk motions.

The spread metric is presented first for dark matter in Fig.
\ref{fig:dmonlyspread}, showing the probability density distribution of the
spread $S$ for dark matter particles either inside (blue) or outside (purple)
of virialized haloes at $z=0$. This quantifies the redistribution of the dark
matter due to any gravitational effects. We see here that the largest spread
distances are significantly larger than any of the characteristic distances
shown in this figure; this is even compared to the largest separation for any
two particles at $z=0$, implying that these distances are much further than
can be achieved from Hubble expansion in voids alone. The overall
distribution follows an exponential decay, with exponentially fewer particles
(once outside the inner $\sim 0.5 \hmpc{}$) being found at larger distances.
There are many possible explanations for these results, from tidal stripping
of objects that end up never merging, accretion of dark matter from
satellites \citep[see e.g. the effects in ][]{VandenBosch2018}, or even
particles on randomised orbits from recently accreted material that end up on
opposite sides of the `splashback' region \citep{Diemer2014, Adhikari2014}.
This splashback region is sometimes larger than the virial radius of the
halo, meaning that two particles may be separated by up to $4R_{\rm vir}$
through this process \citep{Diemer2017a}. Finally, we may expect three-body
interactions between substructures, leading to some being ejected to very
large distances \citep[up to $6R_{\rm vir}$; see][]{Ludlow2009}. This is the
only plausible explanation that we have for such large spread distances in
the dark matter. In practice, we expect the final spread distribution to
reflect the effects of multiple dynamical mechanisms.

In Fig. \ref{fig:dmonlyspread} we also show the consequences of choosing to
average over different numbers of initial neighbours. The simplest metric
would use a single nearest neighbour in the initial conditions. However, the
distance between any two nearest neighbours would be ‘double counted’ and not
representative of motion relative to the surrounding matter distribution in
the case of a single neighbour travelling a long distance. The choice of
$n=3$ is the lowest that ensures that the metric $S$ always represents the
distance between two real pairs of particles, whilst simultaneously solving
this conceptual problem. In practice, the overall distribution of the spread
metric does not depend much on the number of neighbours considered, but we
find that larger choices of $n$ yield a more direct connection between spread
distance and hierarchical structure (with low-spread particles dominating
substructures and high-spread particles corresponding to more diffuse
components, as shown in Fig. \ref{fig:bigdistanceimage}).

\subsection{Baryon Spreading in \simba{}}

\begin{figure}
    \centering
    \includegraphics{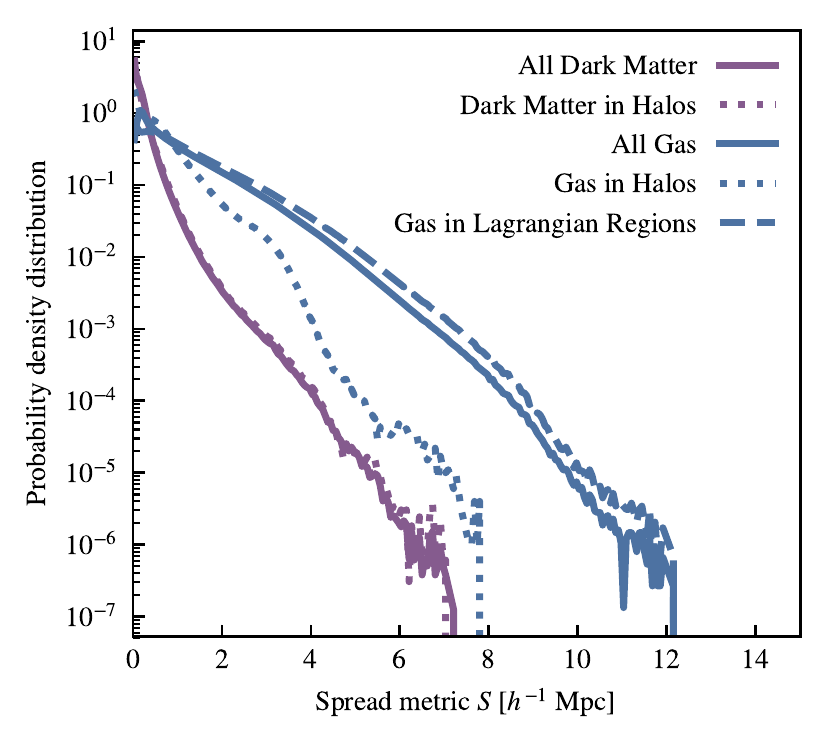}
    \vspace{-0.7cm}
    \caption{Spread distance distribution for gas at $z=0$ (blue) compared to
    that of the dark matter component (purple). Solid lines indicate the full
    distribution, dotted lines correspond to matter inside $z=0$ haloes, and
    the blue dashed line shows the distribution for gas that was inside of
    Lagrangian regions at $z=99$. The distributions for gas inside haloes and
    outside haloes are significantly different, with gas that resides outside
    haloes being preferentially spread to larger distances than gas on
    average. Note that only 10\% of the gas in the entire simulation is in
    haloes at $z=0$. Gas that originated in Lagrangian regions is
    preferentially spread the most, with a factor of 2 offset over the
    unbiased selection at large spread distances.}
    \label{fig:distbaryon}
\end{figure}

Fig. \ref{fig:distbaryon} shows how the distribution of spread distances
for the gas particles is significantly different to that for the dark matter.
Gas particles are able to spread to much larger distances, up to $12\hmpc{}$
(approximately 10 times the virial radius of the largest halo in the box!),
compared to the $7\hmpc{}$ that dark matter can reach. We also see that even
gas inside of haloes at $z=0$ has spread significantly more than the dark
matter when explicitly selecting for this component. This suggests a
different origin for the gas and dark matter content of haloes.

Another interesting component is the gas that originated in Lagrangian
regions (i.e. next to the dark matter that will reside in haloes at $z=0$),
indicated by the blue dashed line. With the baryon fraction of haloes being
typically less than $50\%$ of the cosmic mean, we should expect that a
significant amount of Lagrangian gas is lost over time, possibly spreading to
large distances out of haloes due to high energy feedback events, either
through galactic winds or AGN feedback. In \simba{}, we see that gas from
Lagrangian regions indeed spreads systematically further, with a factor of
$\sim 2$ more particles at distances larger than $\sim 4 \hmpc{}$ than an
unbiased selection would suggest.

\begin{figure*}
    \centering
    \includegraphics[width=\textwidth]{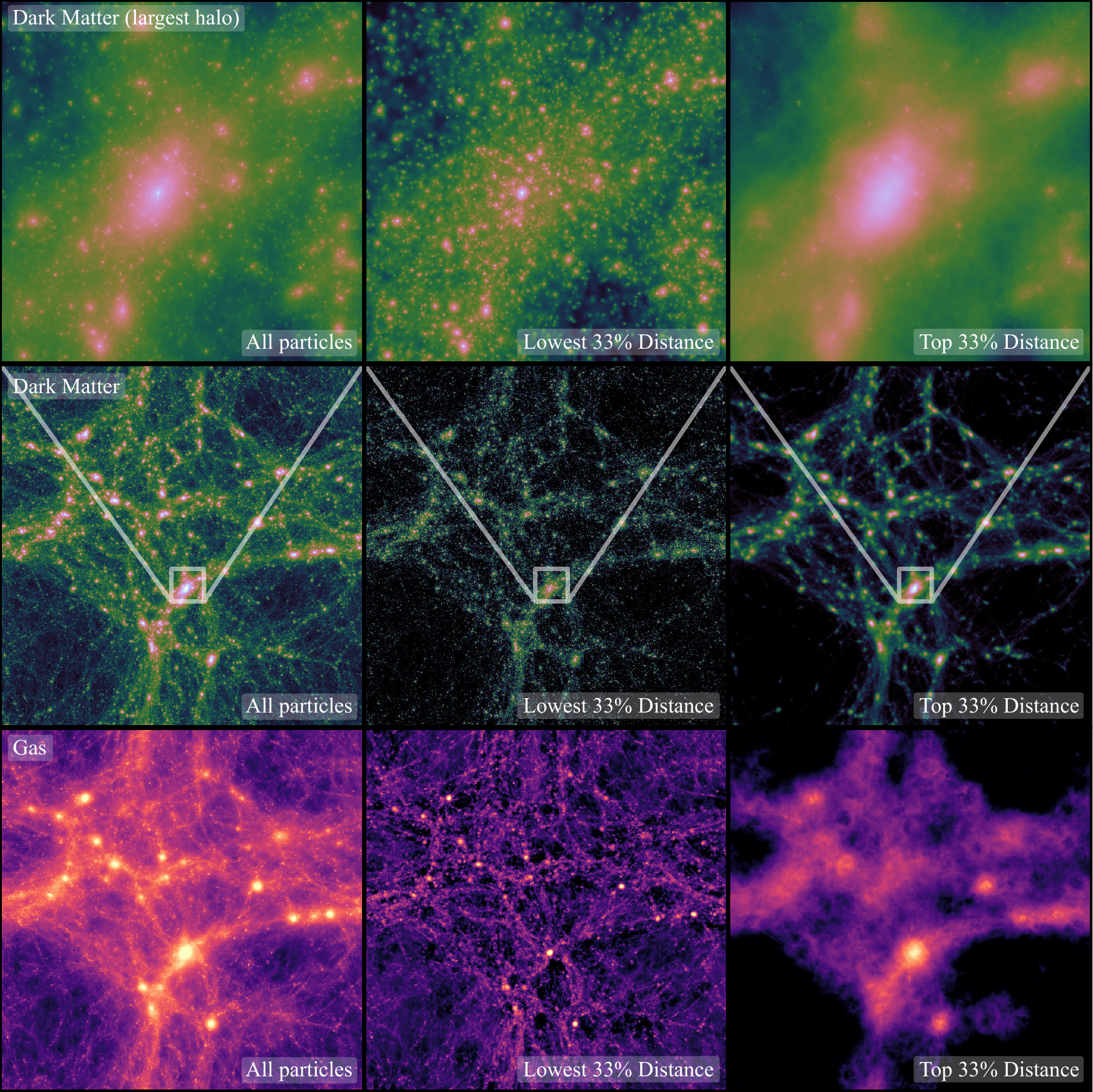}
    \caption{Projected mass surface density distributions for different
    particle selections at $z=0$. The three rows show, from top to bottom,
    the dark matter in a $4.5\hmpc{}$ cubic volume centred around the largest
    halo ($R_{\rm vir}\sim 1.3\hmpc{}$), the dark matter distribution in the
    whole $50\hmpc{}$ box, and gas distribution again in the whole volume.
    Columns show, from left to right, all particles inside of the
    corresponding volume, the 33\% of the particles with the lowest spread
    distance, and the 33\% of the particles that have spread the most. For
    the dark matter, these cuts correspond to particles that have travelled
    less than $0.1\hmpc{}$ and more than $0.25\hmpc{}$, respectively. For the
    gas, these numbers increase to $0.45\hmpc{}$ and $1.25\hmpc{}$,
    respectively, due to the larger spread that gas particles experience.
    Each density projection is generated using smoothing lengths defined to
    encompass the $64$ nearest neighbours and smoothing lengths are kept
    consistent across columns (i.e. they are not recomputed for different
    particle distributions). All density projections in a given row also use
    the exact same (logarithmic) normalisation and colour map to enable
    direct comparisons. Note the significant difference between the spatial
    distribution of material with different spread metric, with sub-structure
    preferentially picked out by the low spread distance selection while the
    large spreads trace large scale structure.
    }
    \vspace{1cm}
    \label{fig:bigdistanceimage}
\end{figure*}

A visualisation of the projected surface densities corresponding to the low-
and high-spread particles is shown in Fig. \ref{fig:bigdistanceimage} for
both dark matter and gas, for the fiducial \simba{} model. We define
`low-spread' particles as those in the lower tertile (33\%) of the
distribution, and `high-spread' particles as those in the upper tertile. By
making these cuts in the distance distribution, we are able to show that the
low-spread particles correspond to substructure, with the high-spread
particles contribution being the larger-scale, more diffuse, CGM and
intergalactic medium (IGM).

\begin{figure*}
    \centering
    \includegraphics[width=\textwidth]{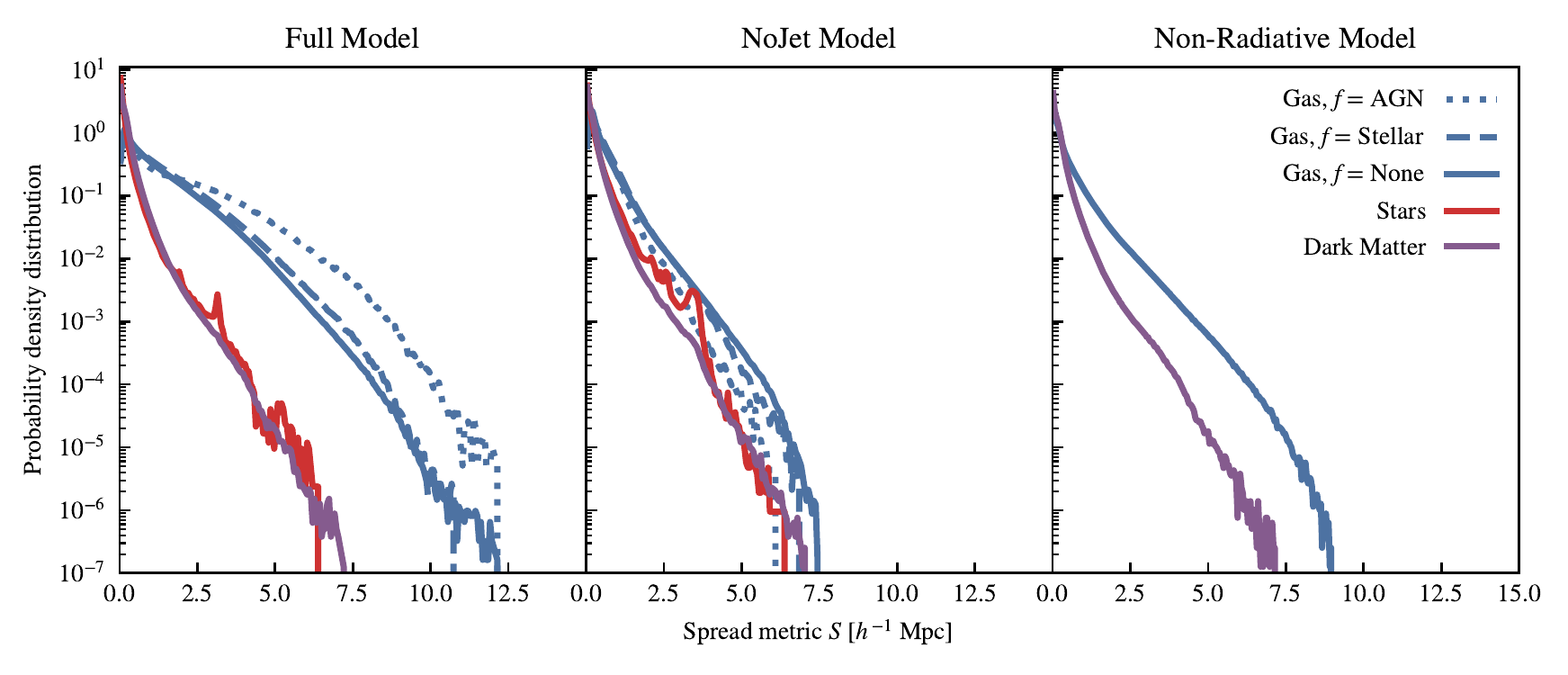}
    \vspace{-0.7cm}
    \caption{Distribution of spread distances split by particle type for gas
    (blue), stars (red), and dark matter (purple). This is shown for the
    $z=0$ particle distribution in the reference model (left), the \nojet{}
    model (center), and the non-radiative simulation (right). The left and
    middle panels separate gas particles that have not been involved in any
    feedback event (solid) from those that have participated directly in
    either stellar (dashed) or AGN (dotted) feedback events. Jets are
    primarily responsible for spreading baryons to the largest distances in
    \simba{}, with significant entrainment of gas that did not directly
    participate in feedback events. The stellar distribution is significantly
    more noisy than the others due to the smaller number of star particles
    (compared to gas or dark matter) in the simulation.
    }\label{fig:feedbackdistance}
\end{figure*}

Considering first the dark matter in the largest halo (top row), we see that
the very small-scale substructure of the halo is preferentially picked up by
the low-spread particles, including the central density peak itself and the
centers of subhaloes. In contrast, the more diffuse dark matter component that
fills the space between these individual density peaks is significantly more
prominent in the high-spread particles, with only a small amount of residual
sub-structure remaining. These trends are also clear at larger scales, as
shown by the view of the $50\hmpc{}$ box in the second row, with large-scale
dark matter filaments primarily traced by high-spread particles. It is
interesting to note that a large amount of structure in voids is not present
in either of these panels, with it being captured by the medium-spread
particles with values $0.1\hmpc{} < S < 0.25 \hmpc{}$. The spread
metric is thus a very useful tool to connect hierarchical structure and
dynamical evolution in cosmological N-body simulations.

The bottom row in Fig. \ref{fig:bigdistanceimage} shows the large scale gas
distributions separated with the same proportions, with a third of the total
gas mass contained in each of the middle and right panels (this corresponds
to different absolute values of the spread metric compared to the dark matter
panels). The low-spread particles trace the densest gas in haloes along with
lower density gas in the central parts of large scale filaments. Of
particular interest is the high-spread gas, which traces the large bubbles
around the most massive haloes that strong AGN jets produce in the \simba{}
model (see \S \ref{sec:fullmodelfeedback}). As expected from Fig.
\ref{fig:distbaryon}, the top third of the gas distribution has been pushed out
to significantly larger distances compared to the third of the dark matter that
moved the most due to gravitational dynamics only. The spread metric hence 
captures the impact of feedback in a global sense.

\subsection{Connecting feedback and the spread of baryons}
\label{sec:fullmodelfeedback}

The kinetic feedback scheme used in \simba{} for both star formation and AGN
feedback makes it straightforward to identify the gas elements that have been
directly impacted by feedback. However, these gas elements will then go on to
entrain and deposit energy into other gas elements as they travel. This makes
it challenging to fully capture the impact of feedback solely from particle
tagging. Here, we use the additional \nojet{} and non-radiative simulations
in order to explore how baryon redistribution is sensitive to different
physics modules in \simba{}, although we caution that these are not fully
independent sub-grid models with their own calibration process.


The left panel of Fig. \ref{fig:feedbackdistance} shows the spread distribution
for the full \simba{} model, splitting the gas component into particles that
have been affected by different types of feedback. Here, AGN feedback takes
precedence over stellar feedback, such that if a particle has been affected
by both it is only classified as being part of the $f=$ AGN group. We see that
the particles that have directly interacted with the AGN are spread to
significantly larger distances, with a vertical offset of 0.5-1 dex compared
to no-feedback particles for $S \gtrsim 5\hmpc{}$. Particles that have been
directly kicked by stellar feedback also have systematically higher spread
metric values, albeit with a smaller offset. This implies that particles are
indeed being spread to these large distances by feedback events.

The left panel also now includes the stellar component, which shows a very
similar distribution to that of dark matter. This is somewhat surprising
given that stars form out of the most bound gas at the center of haloes. It
would be unlikely for a star particle to form from a gas particle with a high
spread value, as these must have been separated dynamically from their
closest dark matter neighbour requiring some form of strong energy injection.
This would eject and heat the particle making it less likely to cool down,
accrete back onto the galaxy, and condense to high enough density to form a
star by redshift $z = 0$. This suggests that the stellar spread distribution
is produced by dynamical effects after the star has formed, affected by the
same physics that shapes the spread distribution for the dark matter,
including tidal disruption and stripping of satellites, merger events, and
orbital divergence through N-body dynamics.

The middle panel of Fig.~\ref{fig:feedbackdistance} shows the spread
distribution for the \nojet{} simulation, where we still include AGN feedback
in the form of radiative winds and X-ray heating but the high velocity jet
feedback mode is disabled. With this change, the spread metric is
significantly affected, with much less difference between the distributions
of the dark matter, gas, and stellar components. While galactic winds and AGN
feedback in radiative mode can still decouple the dark matter and gas
components, high-velocity jets are clearly the dominant mechanism responsible
for spreading baryons to the largest distances in \simba{}. Surprisingly, gas
particles directly kicked by feedback in this case show a lower spread
distribution compared to gas not directly impacted by feedback, in contrast
to the trend seen for the fiducial \simba{} model. This suggests that
feedback in the \nojet{} simulation is not strong enough to compensate for the
fact that feedback events occur in the densest regions (inside galaxies). It is
intrinsically more difficult to escape these deep potential wells, especially
now that a crucial energy injection mechanism from the AGN jets is missing.

\begin{figure}
    \centering
    \includegraphics{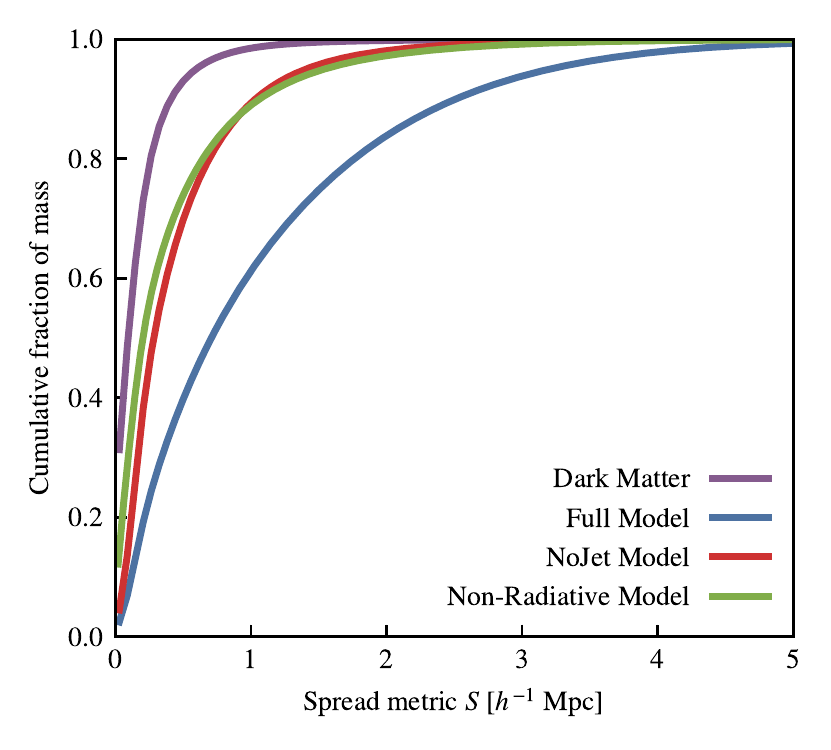}
    \vspace{-0.7cm}
    \caption{Cumulative version of Fig. \ref{fig:feedbackdistance} for the spread of gas
    in the three different models alongside the dark matter from the full model.
    This shows that 10\% of the gaseous matter has spread at least $3\hmpc{}$,
    while 90\% of the dark matter resides within $0.5 \hmpc{}$.}
    \label{fig:cumulativehistogram}
\end{figure}

This result is surprising given that less than $0.4\%$ of gas particles in
the simulation have ever interacted directly with the AGN jets; this has been
enough to significantly decouple the gas from the dark matter dynamically.
Such a high degree of separation points to substantial amounts of gas being
entrained by these powerful jets. It is not simply the case that higher mass
($M_H > 10^{11} \msolar{}$) haloes are quenched internally reducing their star
formation rate; the energetics and dynamics of the CGM and IGM are
significantly altered, as is already seen by the more complex interaction
between the turn-off of the galaxy stellar mass function (GSMF) and the power
of the AGN jets in many studies \citep{Weinberger2018, Dave2019}. 

The final contrast to highlight is the difference between the \nojet{} and
non-radiative model. The non-radiative model shows increased distance between
gas particles and their associated dark matter neighbour compared to the
\nojet{} run; this is due to the lack of cooling preventing particles that
lie in small haloes from remaining as tightly bound. It also highlights how
difficult it is to drive gas into the centers of structures without cooling.
The collisionless dark matter can continue to fall in to bound structures,
with the gas being prevented due to strong accretion shocks. This allows for
a very different kind of separation than what we have shown above for the
full physics model including cooling and feedback.

In Fig. \ref{fig:cumulativehistogram} we show the cumulative version of Fig.
\ref{fig:feedbackdistance} to better show the amounts of mass that are spread
to large distances, showing that 40\% (10\%) of cosmological baryons have
moved > $1\hmpc{}$ ($3\hmpc{}$) by $z = 0$, with a slow tail off ending with
nearly all of the mass being constrained to be spread less than $5\hmpc{}$.

\begin{figure}
    \centering
    \includegraphics[width=\columnwidth]{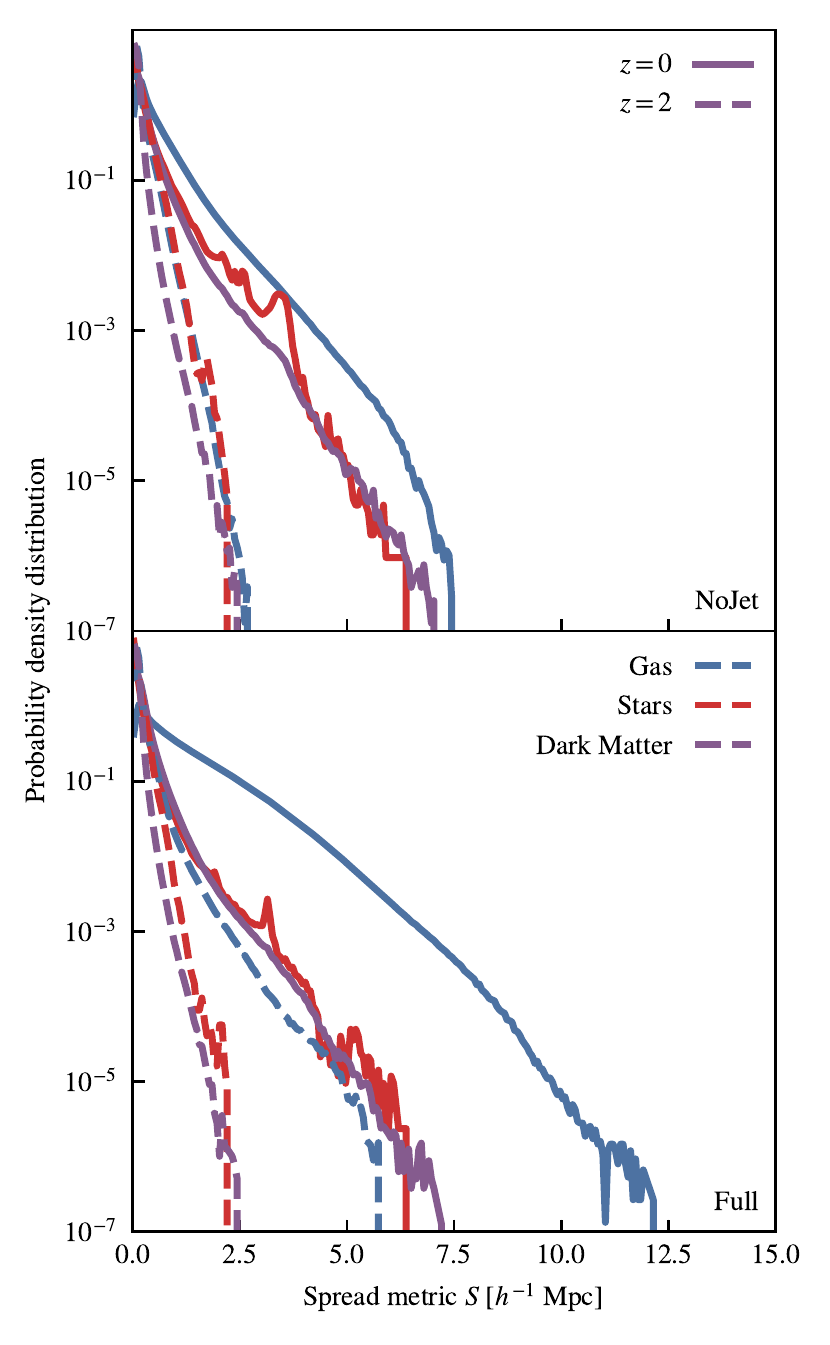}
    \vspace{-0.7cm}
    \caption{Spread metric distributions shown again for the
    \nojet{} (top) and full \simba{} model (bottom) simulations, now including
    the redshift $z=0$ (solid) and $z=2$ (dashed) results. We see that at all
    redshifts the \nojet{} model produces spread metric distributions that
    are highly similar for all three particle types, with the full \simba{}
    model showing divergence between the dark matter and gas even at
    redshift $z=2$. The AGN jets cause a significant difference between
    these gas distributions, and are able to power winds out to a spread of
    $5\hmpc{}$ even by $z=2$.}
    \label{fig:zevodist}
\end{figure}

\subsection{Redshift evolution of the Spread Metric}

From Fig. \ref{fig:feedbackdistance} it is clear that the AGN jets have a
significant impact on the spread metric, causing the maximal spread distance
in the gas to almost double. In Fig. \ref{fig:zevodist} we explore how this
deviation between gas and dark matter depends on redshift. The dashed lines
show the spread metric distribution at $z=2$, and from this we see that in
the full model gas has spread to over $5\hmpc{}$ (more than twice that of the
largest dark matter spread) even by this early epoch. The \nojet{} model
shows no such behaviour, showing a very close convergence between the spread
metrics of all three particle types. This long-distance baryon spreading is
then not a late-time effect; it occurs at all times that the jets are active,
gradually filling in the final spread metric distribution.
\begin{figure*}
	\centering
	\vspace{0.5cm}
	\includegraphics{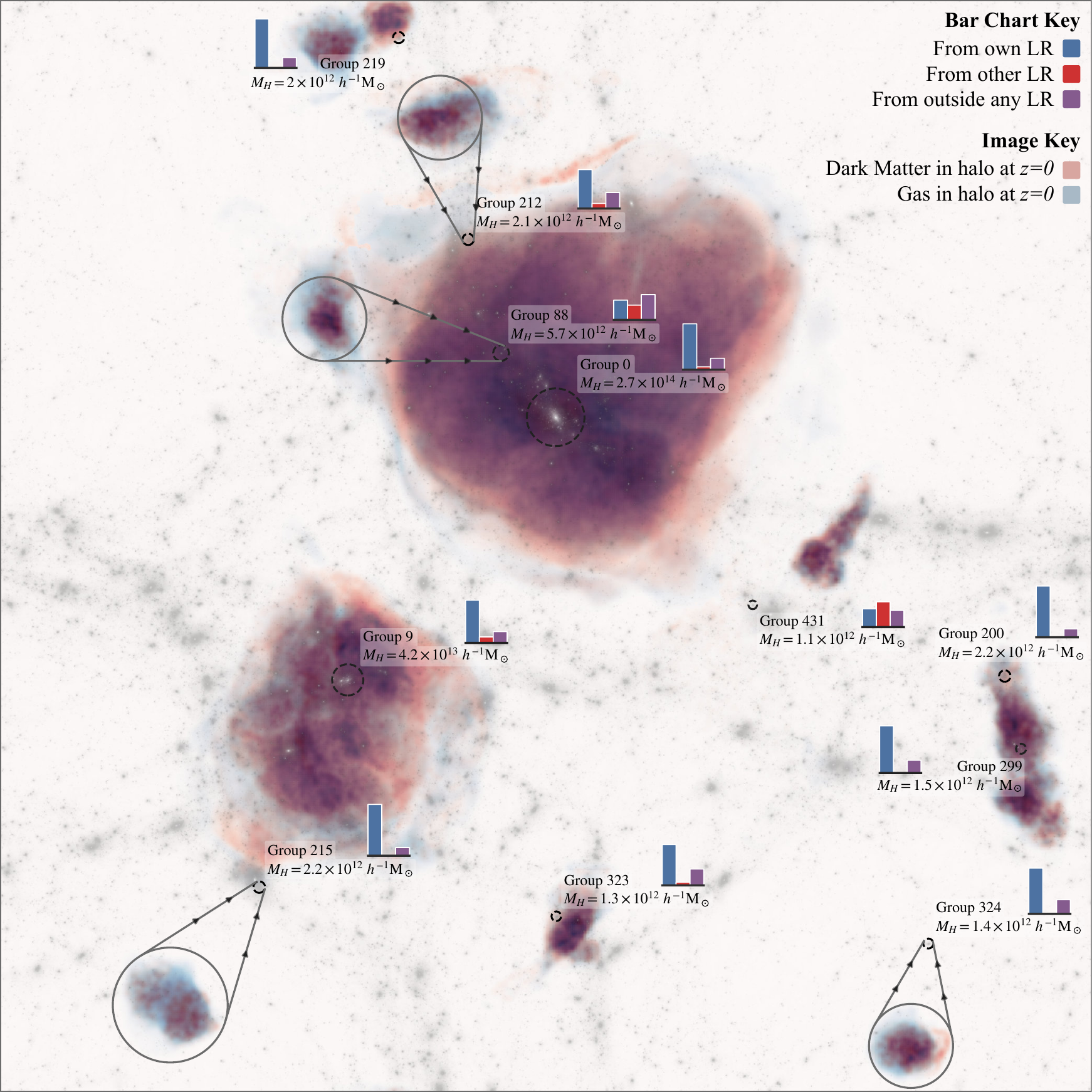}
 \caption{ This visualisation shows two epochs at once, simultaneously
 showing the initial conditions (in blue and red) and the final simulation
 volume at redshift $z=0$ in white/grey. The blue and red show the positions
 of the gas and dark matter (respectively) \emph{in the initial conditions}
 for particles that reside in selected haloes at redshift $z=0$. The overlaid
 white/grey map shows the dark matter at redshift $z=0$ to enable comparisons
 between the initial and final comoving positions for various bound
 structures. For each selected halo, the dashed black circles show their
 virial radii as defined in \S \ref{sec:simba}. For some haloes in crowded
 regions, we have overlaid a circle and arrows showing which blob of dark
 matter and gas in the initial conditions collapses to form this halo.
 Finally, for each halo we show a small bar chart showing how their gas is
 composed from Lagrangian components, as described later in the text. The
 blue bar shows the fraction of gas in each halo that originated from that
 haloes own Lagrangian region, the red bar shows the gas from another haloes
 Lagrangian region, and the purple bar shows the fraction of gas that
 originated outside any Lagrangian region. This figure illustrates the
 significant differences in origin between the gas (blue) and dark matter
 (red) for these selected haloes of various masses. We also see how the
 environment of each halo changes its Lagrangian make-up. In particular,
 group 431 shows a large baryonic component originating from the Lagrangian
 region of another halo, with this halo entering a small cluster environment
 near the end of the simulation. Note that individual regions are
 colour-mapped separately, i.e. the intensity of colour for a single halo is
 unique to that halo only, as to enable all Lagrangian regions to be seen.
 Without this choice, the structure for the lower mass haloes would be
 completely washed out.}
	\vspace{1cm}
	\label{fig:bigtransferpic}
\end{figure*}

\begin{figure*}
    \centering
    \vspace{1cm}
    \includegraphics{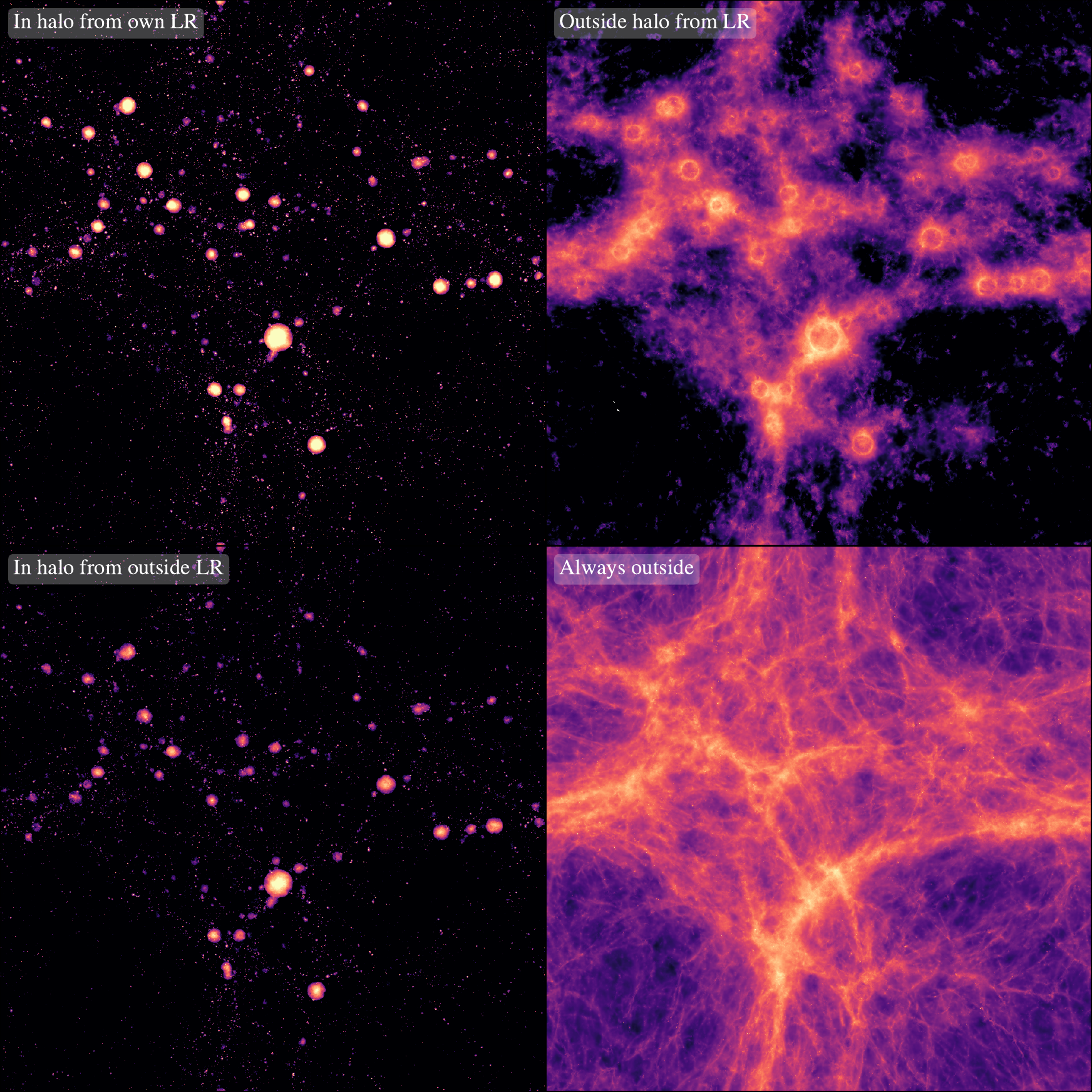}
    \caption{Gas distribution in the fiducial \simba{} model for the full
    $50\hmpc{}$ volume, split by the following Lagrangian components
    (clockwise, starting from top left): particles that began in Lagrangian
    regions at $z=99$ and have remained in the associated haloes at $z=0$;
    particles that began in Lagrangian regions and ended up outside of the
    destination halo; particles that began outside any Lagrangian region and
    ended up outside any halo; and particles that ended up in a halo but
    originated outside any Lagrangian region. All images are shown with the
    same (logarithmic) colour-map and normalisation and taking their linear
    sum would reproduce the full gas distribution at $z=0$. Gas particles
    that began in Lagrangian regions but ended up outside of haloes (top
    right) show a striking similarity to the distribution of gas with the
    33\% highest spread distance shown in Fig.~\ref{fig:bigdistanceimage}.
    As expected, particles that began outside of Lagrangian regions and
    remained outside of haloes (bottom right) trace the filaments and voids.}
    \vspace{1cm}
    \label{fig:lrtransfer}
\end{figure*}

\section{Lagrangian baryon transfer}
\label{sec:transfer}

We have explored the relative motion of dark matter and baryons using a
particle-level metric, showing that AGN jets in the \simba{} cosmological
simulations can spread baryons up to $12\hmpc{}$ relative to the neighbouring
dark matter. In this section, we consider the movement of baryons relative to
dark matter haloes and their corresponding Lagrangian regions. The definitions of
haloes and Lagrangian regions used here are described in \S \ref{sec:simba}.

This topic has been considered recently by \citet{Liao2017}, where they used
a $10\hmpc{}$ non-radiative simulation to show that the gas in haloes may
originate from different places than the dark matter in those same haloes in
the initial conditions.

\subsection{The different origins of baryons and dark matter in haloes}

Fig. \ref{fig:bigtransferpic} illustrates the mixed origins of the gas and
dark matter components in bound structures at $z=0$ by showing simultaneously
the initial and final states of the simulation. A common trend for all haloes
is a shell of gas around the main dark matter component in the initial
conditions, showing that gas in general is able to collapse further (due to
cooling and other processes) than the dark matter, which is unable to lose
angular momentum as efficiently. This is consistent with the larger values of
the spread metric for gas in haloes relative to the dark matter in haloes, as
shown in Fig. \ref{fig:distbaryon}.

The origin of the dark matter in the initial conditions corresponds exactly
to our definition of Lagrangian region for that component in \S
\ref{sec:simba}. These Lagrangian regions have very complex shapes, with
larger haloes tending to have more spherical Lagrangian regions, as can be
seen with the largest halo in the box (Group 0) in Fig.
\ref{fig:bigtransferpic}. These complex non-spherical shapes are why we
chose to identify our Lagrangian regions for gas through neighbour searching,
as other methods (e.g. constructing a convex hull enclosing all dark matter
particles that end up in a given halo) would not allow us to capture the
surprisingly intricate structure that is at play here.

There are many possible reasons for the complex shapes that we see here.
Consider a simple case where we have one `main' halo, and a satellite that is
being accreted. The gas and dark matter in the satellite galaxy have several
potential fates. For instance, when accreting onto the main halo, the gas in
the satellite may be shock heated, and stalled in the CGM, with the dark
matter being able to continue to move towards the center of the main halo.
This process dynamically separates the dark matter and gas, and now the gas
may have several fates; it could be pushed out in a feedback event, rise out
of the halo due to buoyancy, or fall to the centre of the halo after cooling
and re-join the dark matter. Once the gas has been removed from the CGM into
the IGM, it is free to be picked up by other passing galaxies.

The other possibility for the fate of this substructure is the dark matter
failing to accrete onto the central. In this case, the dark matter continues
moving out into the IGM, with the gas being shocked and captured by the main
halo. It is this complex difference in assembly between dark matter and
baryons, due to the latter behaving as a collisional fluid, that we aim to
capture here.

\subsection{Computing transfer between Lagrangian regions}

Given the definitions of haloes and Lagrangian regions in
\S \ref{sec:simba}, it is possible to classify every particle in the
simulation according to their Lagrangian ID and halo ID (if any) in the
initial and final conditions. The algorithm is as follows:

\begin{enumerate}
	\item ID match all particles between the initial and final conditions, including
	      star particles (these are matched to their gas progenitor). Black holes are ignored in this analysis since globally they represent a minimal amount of mass.

	\item Every particle at $z=0$ has several possible final states and origins, based on its halo ID ($i$) and Lagrangian region ID ($j$):
	      \begin{itemize}
	            \item Particle resides in  halo ($i \neq -1$)
	            \begin{itemize}
	           		\item Particle originated in the same Lagrangian region, $j = i$
	           		\item Particle originated outside any Lagrangian region, $j \equiv -1$
	           		\item Particle originated in some other Lagrangian region, $j \neq i$
	            \end{itemize}
	            \item Particle resides outside of any halo ($i \equiv -1$)
	            \begin{itemize}
	            	\item Particle originated outside any Lagrangian region, $j = i$
	            	\item Particle originated in some Lagrangian region, $j \neq i$
	            \end{itemize}
	      \end{itemize}
	      
	\item For every halo and Lagrangian region, the mass originating from each
	      of the above components is computed and stored.
\end{enumerate}

A visualisation of this particle classification scheme is shown in
Fig.~\ref{fig:lrtransfer}, where we split the gas distribution in the
\simba{} $50\hmpc{}$ box into the four main Lagrangian components that we
consider in the remainder of this paper. Considering each panel clockwise
from the top left, we select first the gas that is in the same halo at
redshift $z=0$ as the Lagrangian region that it originated in. As expected,
we see a population of spherical shapes corresponding to every halo in the
box, with their sizes corresponding to $R_{\rm vir}$ as defined by AHF. The
centers of haloes, where the gas is densest, are the brightest.

In the top right panel we have the gas that is outside any halo at $z=0$, but
is assigned to a Lagrangian region at $z=99$; this is the gas that should
have ended up in haloes by the end of the simulation if the baryonic matter
was also collisionless. We see that this component traces gas primarily
around massive haloes, resembling the large-scale bubbles that the AGN jets
power in \simba{} \citep{Dave2019}. Note that some of this gas piles up just
outside of haloes due to the somewhat arbitrary boundary defined by the
virial radius of haloes. This gas resides primarily in filaments, with some
reaching out into the voids.

In the bottom right panel, we visualise the gas that begun outside any
Lagrangian region and resides outside any halo at redshift $z=0$. This gas
traces the majority of the filamentary structure, and shows all of the
structure in the voids. 

Finally, in the bottom left panel, we have the gas that is in haloes at $z=0$
but originated from outside any Lagrangian region. As expected, this shows a
very similar structure (albeit less bright) to the gas that resides in its
own halo (top left), but this component originates from regions where the
dark matter now resides outside of haloes. This gas is likely dragged into
these bound structures by cooling flows, while the dark matter
is not able to lose angular momentum quickly enough to assemble by $z=0$.

\subsection{Transfer in a non-radiative Model}

\begin{figure}
	\centering
	\includegraphics[width=\columnwidth]{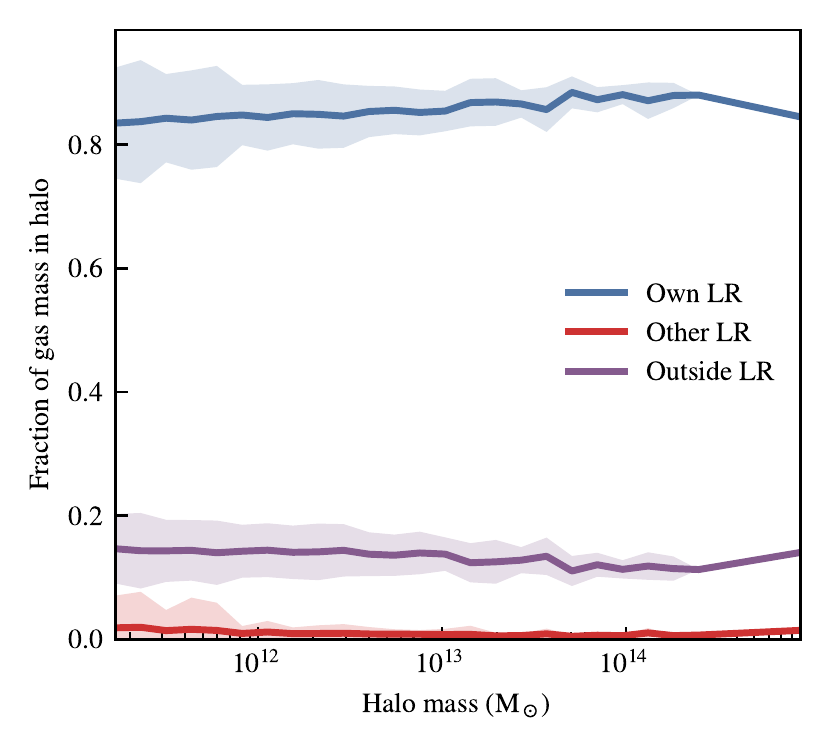}
	\vspace{-0.7cm}
 \caption{ The fraction of baryonic mass originating from each Lagrangian
 component in the non-radiative model (i.e. without sub-grid physics) is
 shown as a function of redshift $z=0$ halo mass. The gas particles are
 binned by their origin, with the baryons originating from their own
 Lagrangian region shown in blue, the Lagrangian region of other haloes
 (red), and outside of any Lagrangian region (purple). Shaded regions show
 the $1\sigma$ scatter in a given bin, which is given by one standard
 deviation of variation. The lines represent the mean value within each bin.
 Approximately 85\% of the baryonic mass of a given halo originates from its
 own Lagrangian region, showing very little transfer of baryons from either
 outside or from another Lagrangian region. This is provided for comparison
 to the full model result in Fig. \ref{fig:maintransferresult}.}
	\label{fig:nonradiativetransfer}
\end{figure}

Before considering the numerical results of the full model, we first present
the non-radiative simulation as a null model to investigate the effects of
hydrodynamics alone. In this case, we run the simulation without cooling,
star formation, or feedback, only including hydrodynamics, cosmology, and
gravity. In Fig. \ref{fig:nonradiativetransfer} we present the fraction of
baryonic mass for each halo contributed from each Lagrangian component, as a
function of halo mass. The blue line shows the fraction of mass in each halo
from its own Lagrangian region (top left in Fig. \ref{fig:lrtransfer}), the
red shows transfer into a halo from another Lagrangian region, and the purple
line shows the fraction of baryonic mass from outside any Lagrangian region
(bottom left in Fig. \ref{fig:lrtransfer}). There is no dependence on halo
mass (as the simulation is effectively scale-free above some resolution
limit), and apart from some small level of transfer from outside any
Lagrangian region (of around $10-15\%$), the baryonic mass in each halo
consists of that which originated in its own Lagrangian region.

The difference in origins of the baryons in the final haloes, from
hydrodynamical effects alone, is then around the $10-15\%$ level. This is
close to the 25\% level of segregation between gas and dark matter reported
by \citet{Liao2017} (who also used a non-radiative simulation), with the
difference likely rooted in the definitions that we use. We consider the
fraction of gas particles in the final redshift $z=0$ halo whose initially
pairing dark matter is also resident in that halo; hence what we are really
counting is the `contamination' of the halo by gas particles from outside of
its Lagrangian region. \citet{Liao2017} count all particles in the final
halo, treating gas and dark matter equally, then finding all particles that
were gas-dark matter pairs in the initial conditions. Their higher level of
segregation is expected due to contributions from dark matter particles that
are resident in a halo but whose initial gas pair is not. Fundamentally this
represents the difference in our approaches; here we are interested in
treating the dark matter as a ground source of truth, and asking if the gas
nearest to that dark matter follows it into the same haloes.
\citet{Liao2017}, on the other hand, were interested in treating \emph{all}
occupants of the final halo as the ground source of truth, and asking what
differences there were in their origin.

The causes for our contamination here are less clear than in the case of
\citet{Liao2017}; we would report a halo that has had gas only \emph{removed}
as being completely uncontaminated, and hence stripping of gas is
an unsatisfactory explanation of these differences. The likeliest explanation
for the contamination in this case is that the baryons and dark matter go through
a phase of mixing as they enter the cosmic web, before going on to fully
collapse into bound structures.

\subsection{Transfer \emph{into} haloes}
\label{sec:transferinto}

\begin{figure*}
	\centering
	\includegraphics{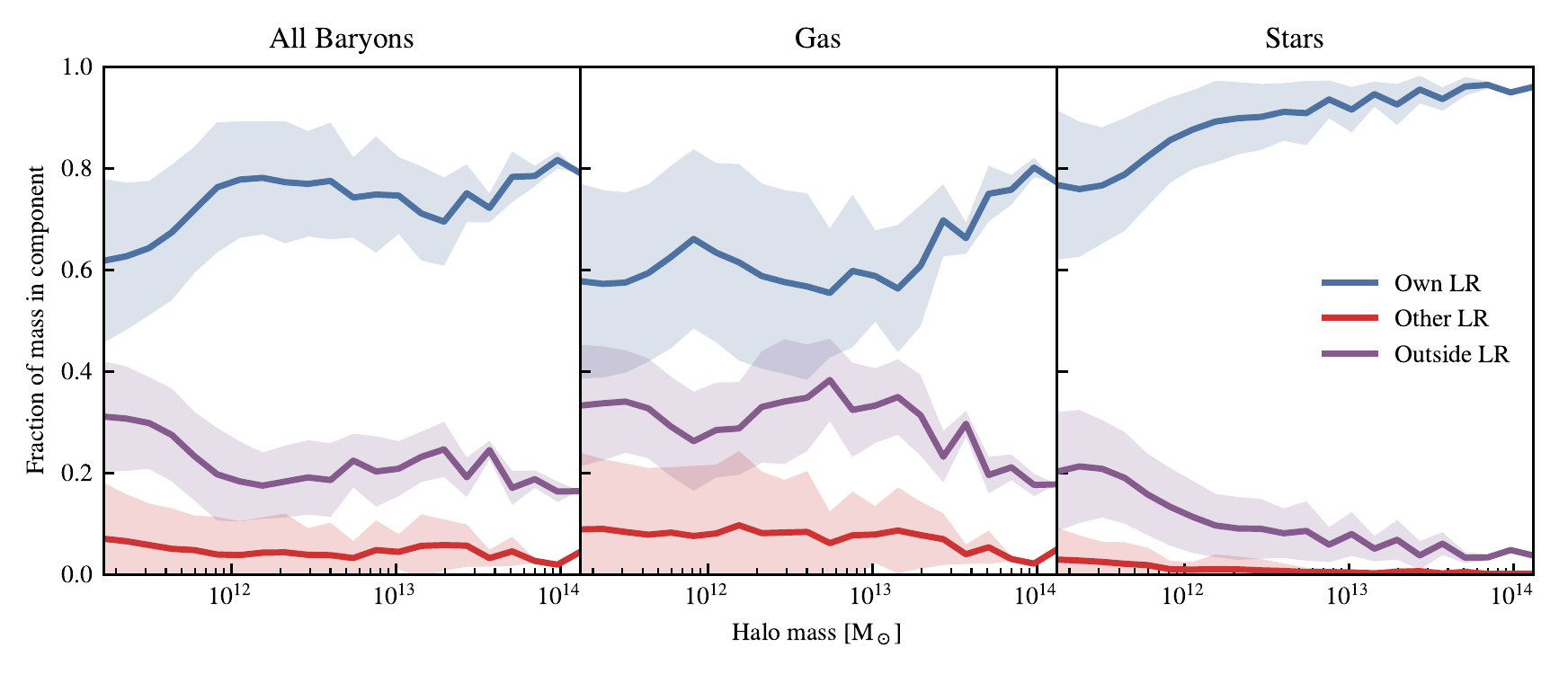}
	\vspace{-0.7cm}
	\caption{
  The fraction of baryonic mass in haloes at $z=0$ originating from their own
  Lagrangian region (blue), the Lagrangian region of other haloes (red), and
  outside of any Lagrangian region (purple), shown as a function of $z=0$
  halo mass for the fiducial \simba{} model. We consider all baryons in
  haloes (left) as well as their gas (center) and stellar (right) components
  separately.
	}
	\label{fig:maintransferresult}
\end{figure*}

Moving on to the full \simba{} model, we consider again the fractions of
baryonic mass as a function of halo mass, split by Lagrangian component. Fig.
\ref{fig:maintransferresult} shows three panels: the left panel shows all
baryons, the centre shows only gas, and the right panel shows the
contribution from only the stars. The lines are coloured the same as the
non-radiative model shown in Fig. \ref{fig:nonradiativetransfer}. Now that we
have introduced scale into the simulation through density-dependent energy
injection mechanisms, these components scale with halo mass. The general
trend is that for an increasing halo mass, a Lagrangian region is able to
hold on to more of the original baryonic mass, with this flattening off
around $M_H = 10^{12} \msolar$. For a given halo, significantly more of the
gaseous mass originates outside the original Lagrangian region as compared to
the stellar mass ($\sim 40 \%$ versus $\sim 10 \%$). The transfer between
haloes is at around the $\sim 10\%$ baryonic mass level, with this transfer
predominantly originating from the gaseous component, as compared to the
stellar component. This combines nicely with the distance metrics shown in \S
\ref{sec:feedbackmetrics}, which showed that the dark matter and stars have
very similar dynamics and hence should be similarly well bound.

This transfer into, and between, Lagrangian regions can have several physical
origins. The first, as shown in the non-radiative run, is caused by the
collisional dynamics of the gas preventing gas from following the dark matter
in all cases. We found that this can account for up to 15\% of the baryonic
mass of a bound structure at redshift $z=0$ originating from a different
region than the dark matter (see Fig. \ref{fig:nonradiativetransfer}), but this
could not account for any \emph{inter-Lagrangian} region transfer.

The galaxy formation sub-grid model clearly has a significant effect on the
baryonic make-up of haloes at redshift $z=0$. The fraction of mass from
outside any Lagrangian region has increased to 20-40\%. This increase is
explained by the inclusion of sub-grid cooling and feedback processes, with
the baryons now able to cool before accreting and lose angular momentum at a
much higher rate than the dark matter component is able to.

Around 10\% of the baryonic mass of haloes is now made up of gas that has
experienced inter-Lagrangian transfer. It is important to recall that this is transfer
between bound structures at redshift $z=0$, and that it only takes into account
the initial and final conditions of the simulation; we do not know the complete history
of these particles.

The transfer between haloes has several possible sources: stripped gas from
nearby galaxies that are still classified as their own bound structures at
redshift $z=0$, gas that has been expelled from galaxies through stellar
winds or AGN feedback and re-captured by a halo, and transfer due to boundary
effects caused by the complex shapes of Lagrangian regions according to the
definition adopted. With the non-radiative simulation showing zero transfer
between haloes, and there being little transfer before $z=2$ in the fiducial
model (see below in Fig. \ref{fig:ltzevo}), we believe that the contribution
from pure dynamics alone to inter-Lagrangian transfer is likely very small.
When repeating this analysis with the \nojet{} run, the inter-Lagrangian
transfer is reduced, but still remains at the 10\% level. The feedback events
that power this transfer must be dominated by the expulsion (or alternatively
preventative pathways) from stellar winds and the residual thermal AGN
feedback.

A given mass bin contains haloes that entertain a range of 10x in transfer,
which is likely dependent on environment. Future work should investigate in more detail
the physical mechanisms driving the scatter in these relations.

The level of transfer above a halo mass of $10^{13} \msolar{}$ must be
interpreted carefully, as there are very few haloes above this mass present
in the box (less than 50), with the small scatter being misleading. It is
also important to note that the shaded regions in Fig.
\ref{fig:maintransferresult} represent the $1\sigma$ scatter in a given bin
and explicitly do \emph{not} include any dispersion that would occur from a
finite sampling of haloes or halo assembly bias.

\subsection{Redshift evolution of transfer into haloes}

\begin{figure}
    \centering
    \includegraphics[width=\columnwidth]{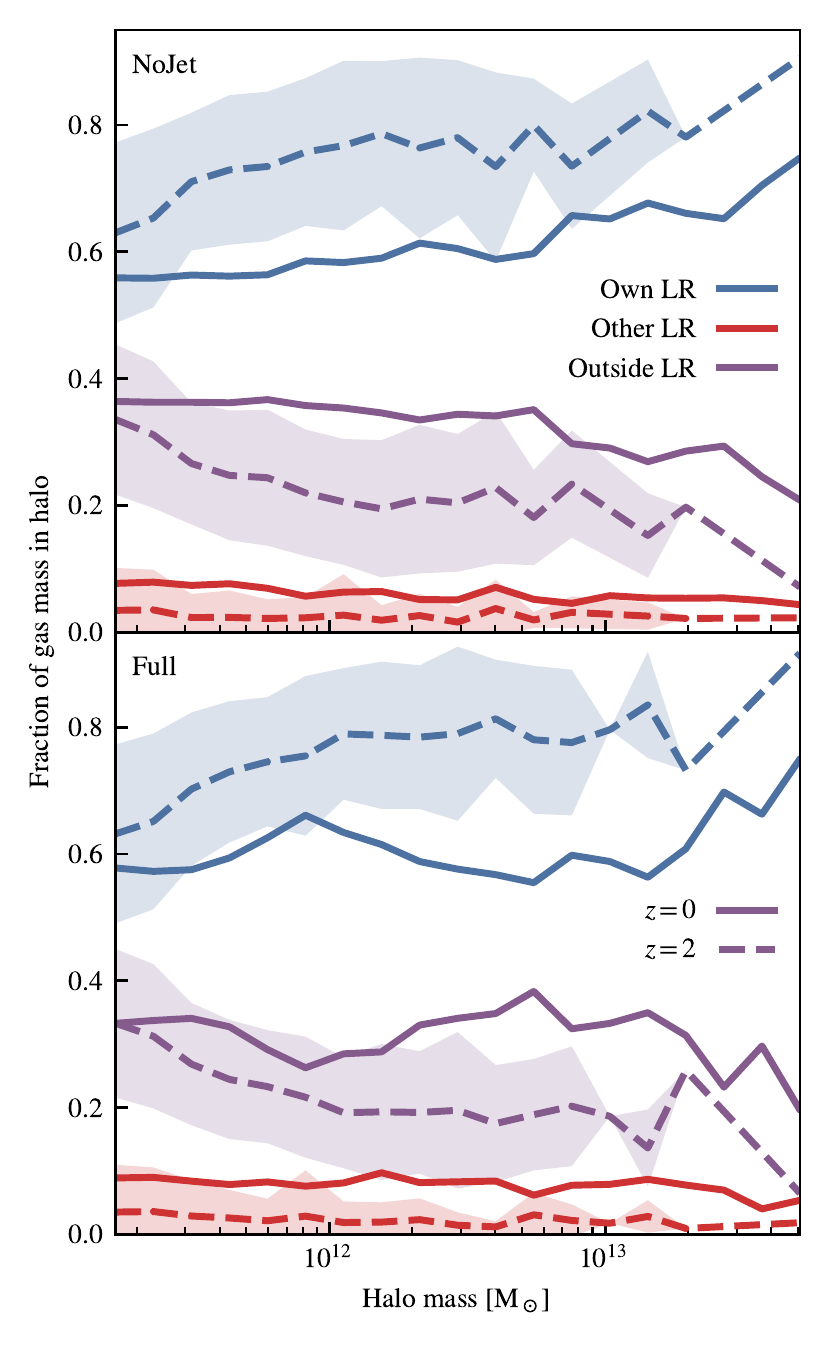}
    \vspace{-0.7cm}
    \caption{The fraction of gas mass in haloes at redshift $z=0$ (solid) and
    redshift $z=2$ (dashed) as a function of halo mass at that redshift, split by
    Lagrangian component. Scatter is shown only for the $z=2$ results. The top panel
    shows the results from the \nojet{} simulation, with the bottom showing the
    full \simba{} model.}
    \label{fig:ltzevo}
\end{figure}

To further investigate the origin of the inter-Lagrangian transfer, in Fig.
\ref{fig:ltzevo}, we consider the \nojet{} model and show how the gas in
haloes at redshift $z=2$ is composed in this and the full \simba{} model.

We see that both the \nojet{} and \simba{} models broadly reproduce the same
fractions of gas in each Lagrangian component, with some interesting differences.
In the full model, a higher fraction ($25-50\%$) of the halo gas originates
from inter-Lagrangian transfer than the \nojet{} model at all masses, with no
change in the shape of this function observed. The fraction of gas
originating outside of any Lagrangian regions shows a dip at around
$10^{12}\msolar{}$ being removed in the \nojet{} model, however this is well
within the scatter that we observe in the full model results.

All of this is despite both models producing very different $z=0$ halo baryon
fractions (see Fig. \ref{fig:baryonfraction} for the full model; the \nojet{}
model produces baryon fractions at approximately the cosmic mean for all halo
masses above $\sim10^{11}\msolar{}$). For a further investigation, halo matching
should be performed between the two models and individual cases compared, but
this is out of the scope of the current work.

The fraction of gas in haloes originating from the different Lagrangian
components shows a closer match at $z=2$, with the shape and
normalisation of all components being well within the reported scatter. The
higher-mass end of these results ($M_H > 10^{13}\msolar{}$) also lacks
objects here, with there being even fewer in this mass range than at $z=0$.

We see that between redshift $z=2$ and $z=0$ a change in the slope
of these functions takes place, and that the level of inter-Lagrangian transfer
increases significantly. The fraction of gas originating from the Lagrangian 
regions of other haloes increases by a factor of two (or more) at all halo
masses, with the fraction of transfer from outside Lagrangian regions remaining
constant or again increasing by a factor of two dependent on the resident halo mass.

All of this must be explained within the context of very different baryon
fractions for all haloes at $z=0$. One possibility is that the majority of
gas gained from outside of a haloes own Lagrangian region remains in the CGM,
with very little of it making it into the disk (this is supported by the very
low fraction of halo stars that originate from transfer, see Fig.
\ref{fig:maintransferresult}). This gas can then be swept out of the halo
either by stellar winds or (ejective) AGN feedback. Alternatively, if the
main pathway for feedback is preventative, and the gas outside of haloes is
well mixed, then this assembly of baryons would be curtailed equally for all
Lagrangian components. A further investigation of these transfer properties
(considering differences between the galaxy disks and the CGM) would be well
suited for follow-up work using higher resolution simulations.

\subsection{Transfer \emph{out} of Lagrangian Regions}

\begin{figure}
	\centering
	\includegraphics{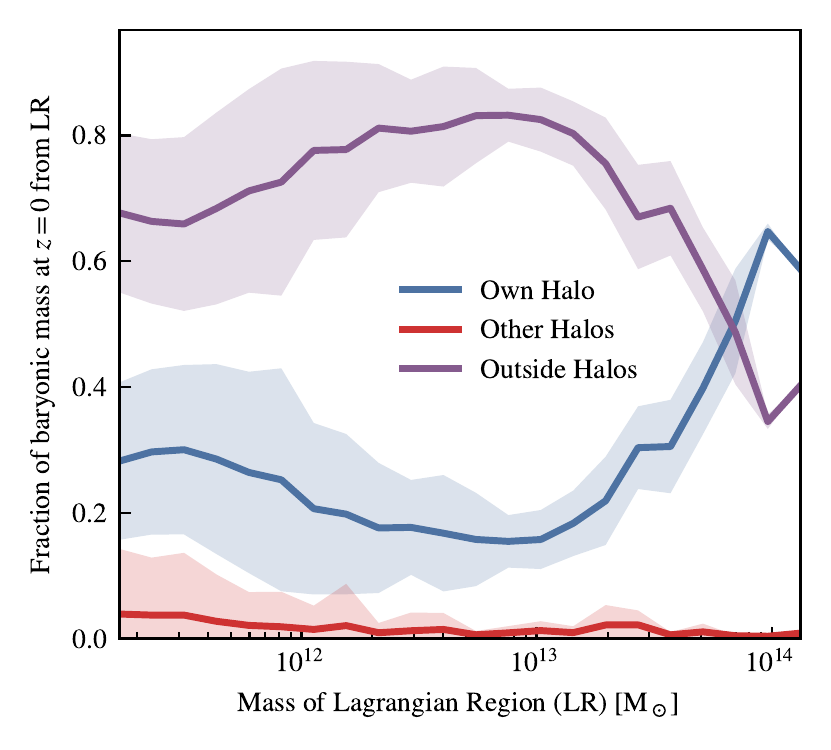}
	\vspace{-0.7cm}
 \caption{The fate of gas that begins in Lagrangian regions, as a function of
 initial Lagrangian region mass. The blue line shows the fraction of baryons
 that reside in the halo that defines the Lagrangian region at redshift
 $z=0$, the red line shows the the fraction of baryons that lie in a
 different halo, and the purple line shows the baryons that lie outside of
 any halo at redshift $z=0$. All but the most massive objects in the box
 struggle to retain more than 30\% of their baryons due to various factors,
 see the text for details. The fraction of mass retained in the corresponding
 halo (blue) is the lowest in the mass range $10^{12} - 10^{13}\msolar{}$.}

	\label{fig:transferoutoflrs}
\end{figure}

Let us now consider the fates of baryons that begin their lives in Lagrangian
regions. This material has three possible fates, as shown in Fig.
\ref{fig:transferoutoflrs}: it can end up in the same halo as the dark matter
from that Lagrangian region (blue line), in another halo (red line), or
outside of any halo in the IGM (purple line). Here, we we plot the fraction
of LR mass at $z=0$ from each component as a function of their Lagrangian
region mass (this is the sum of the baryons and dark matter contained within
that Lagrangian region). The Lagrangian region mass is somewhat higher than
the eventual halo mass due to the baryon fractions of redshift $z=0$ haloes
being below the cosmic mean. We see that, below a halo mass of
$10^{13.5}\msolar{}$, only around 20-30\% of the baryons initially present in
the Lagrangian region make it in to the halo by $z=0$. Only above a halo mass
of $10^{13.5}\msolar{}$ do haloes become strong enough attractors to retain
the majority of their baryons. Despite the clear trend, this result is
somewhat uncertain due to the very small number of these very large haloes
present in our $50\hmpc{}$ box. On top of this initial structure, we see that
there is a dip in the retained fraction of baryons between $10^{12}$ and
$10^{13}\msolar{}$. We speculate that this is due to the increased efficiency
of AGN feedback in haloes in this mass range, allowing for more gas in
central objects to be expelled, however making a direct connection would
require significant investigation. It is worth noting that without the AGN
jets (i.e. in the \nojet{} run), the baryon fraction of haloes in this mass
range is approximately $f_b / f_{b,c} = 1$.

Finally, we find that up to 10\% of the Lagrangian region gas of low-mass
haloes ($<10^{12} \msolar{}$) can be transferred to other haloes, decreasing at
higher masses. A larger cosmological volume with more objects is required for
a full study of objects at masses higher than $M_H > 10^{13}\msolar{}$, but
these trends point towards inter-Lagrangian transfer being fuelled by
accretion of gas that is either expelled or stripped from lower mass haloes
by higher mass objects. A plausible physical scenario is that early
feedback leading up to redshift $z=2$, where star formation (and hence
stellar feedback) peaks, expels significant quantities of gas from lower mass
haloes that can then be swept up at later times from the IGM by all haloes.
Higher mass haloes at this redshift may have a strong enough gravitational
potential to enable their stellar winds to be more efficiently recycled,
preventing them from being sources of inter-Lagrangian transfer.

\begin{figure}
	\centering
	\includegraphics{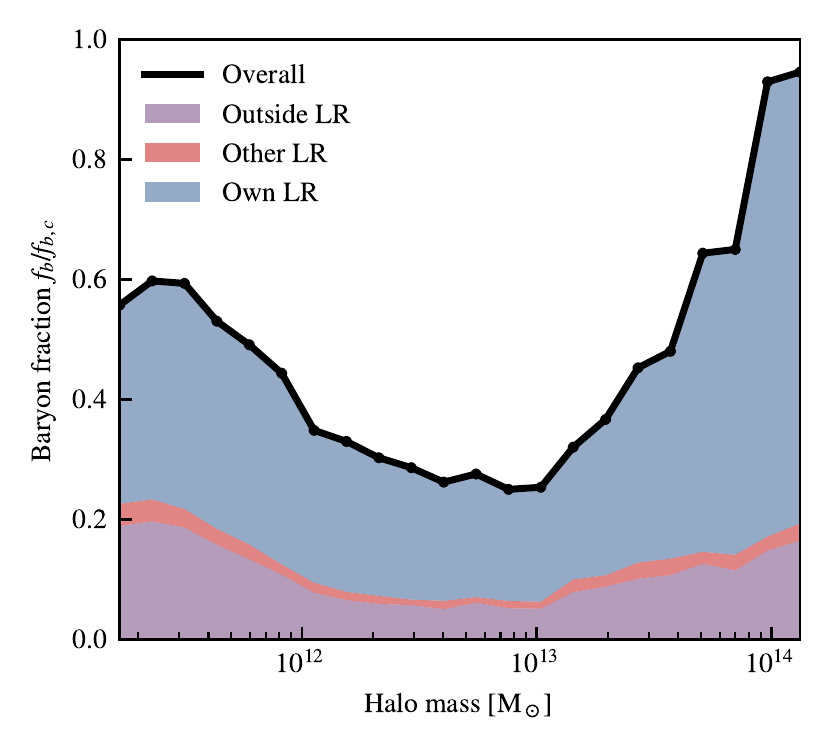}
	\vspace{-0.7cm}
	\caption{The baryon fraction $f_b$ relative to the cosmic baryon fraction
	$f_{b, c}$ shown as a function of halo mass. The coloured bands show the
	contributions to the baryon fraction from various Lagrangian components.}
	\label{fig:baryonfraction}
\end{figure}

The combination of the baryons that are retained by haloes (Fig.
\ref{fig:transferoutoflrs}) and the baryons that they manage to accrete from
sources outside their Lagrangian region (Fig. \ref{fig:maintransferresult})
is seen in the baryon fraction of haloes, shown in Fig.
\ref{fig:baryonfraction} split by Lagrangian component. Here, we split the
overall baryon fraction (relative to the cosmic mean) into three Lagrangian
components, coloured by the baryons from the haloes own Lagrangian region
(blue), other Lagrangian regions (red), and from outside any Lagrangian
region (purple). In general, we see that there is a trough in the baryon
fractions of haloes with a mass between $10^{12}\msolar{}$ and
$10^{13}\msolar{}$, with the baryon fraction reaching the cosmic mean for the
largest objects in the box (with a halo mass of $10^{14}\msolar{}$). The
baryon fraction returning to $f_b = 1$ for these very large haloes is not due
to these haloes retaining all of their Lagrangian gas, however; it is a
complex interplay between their accretion from outside, from other Lagrangian
regions, and from the significant component that originates outside of any
Lagrangian region. These objects are clearly able to mix outside of their
halo boundaries, swapping gas with the IGM, as has been shown in several
studies through `splashback' \citep{Mansfield2017, Diemer2017}.

The dip in baryon fraction between $10^{12}\msolar{}$ and $10^{13}\msolar{}$ in halo
mass corresponds to the dip in retained baryons in a similar mass range in
Fig. \ref{fig:transferoutoflrs}. However, within this mass range, it appears
that the fraction of baryons originating from outside the Lagrangian region is
more significantly affected than the fraction of baryons from the haloes own
Lagrangian region (reduced by 50\% as opposed to 20\%). This points
to a more complex accretion history for these objects, with a mixture of
ejective feedback (in general reducing the amount of retained baryons) and preventative
feedback (in general reducing the amount of baryons from outside of the
corresponding Lagrangian region) shaping their baryonic content.

\section{Variations on numerical parameters}
\label{sec:convergence}

The above halo-based metrics will have a certain level of dependence on the
choice of halo finder used. In an attempt to ensure independence of the
results from such factors, the above analysis was repeated  with the 3D
friends-of-friends (FoF) halo finder included in the {\tt yt} package
\citep{Turk2011}. We also repeated the analysis with the \velociraptor{} 6D
FoF finder \citep{Elahi2019}. The latter will disentangle active mergers, but
as active mergers make up a small fraction of the galaxy population, the
above results are qualitatively unaffected and only change quantitatively
to the 5\% level. The use of a FoF finder, rather than the spherical
overdensity finder found in AHF, did not qualitatively change the results.

In this section, we explore the implications of extending the Lagrangian
region of haloes while retaining the ability to capture non-uniform shapes. We
find that, in general, including more particles in the definition of the Lagrangian
region (than are present in the halo) leads to a fractionally higher level
of inter-Lagrangian transfer and more self-contribution to the final halo mass
at the expense of transfer from outside any Lagrangian region. This is expected, as
now many more particles are classified as being present in the Lagrangian region.

\subsection{Filling in Holes in Lagrangian Regions}

\begin{figure}
	\centering
	\includegraphics{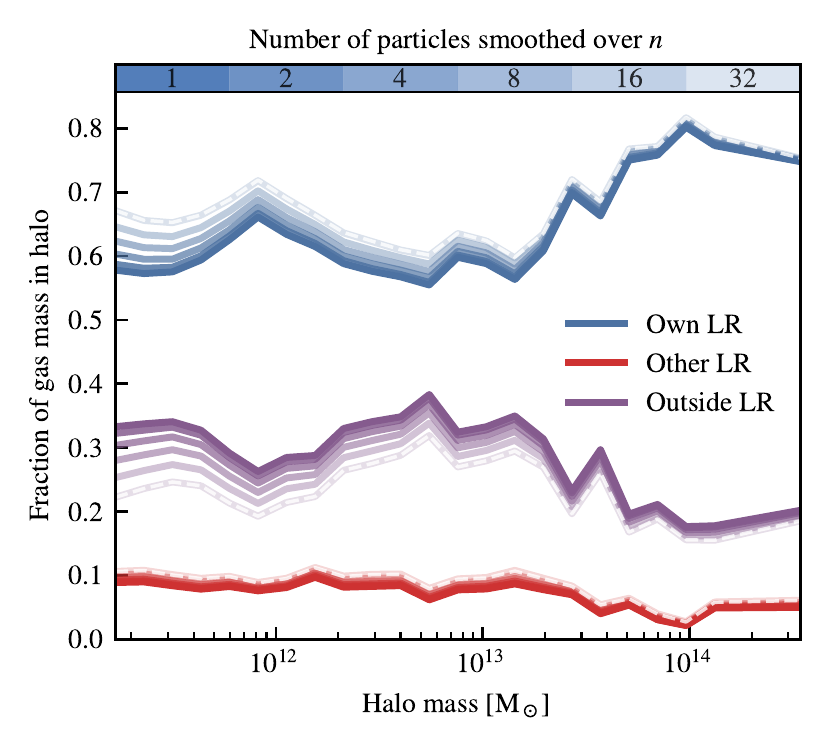}
	\vspace{-0.7cm}
 \caption{The same as Fig. \ref{fig:maintransferresult}, but including
 Lagrangian region smoothing. Each line, coded by transparency, shows the
 fraction of gas mass in a halo from each component when the Lagrangian
 regions have been smoothed by 1 (i.e. the fiducial result), 2, 4, 8, 16, or
 32 particles (from darkest to lightest respectively). The white dashed line
 shows the result for the 32-smoothing case where the particles are given to
 the highest, rather than lowest, mass haloes; no difference is seen here
 suggesting that there is little overlap between the Lagrangian regions on
 these scales. See the text for the details of how this smoothing is
 constructed.}
	\label{fig:smoothconv}
\end{figure}

Our method for producing Lagrangian regions simply uses the dark matter
particles from a given halo; this naturally leads to a very diffuse
Lagrangian region. To see how the diffuse nature of these regions affects our
results, we smooth out the Lagrangian regions, by extending the procedure
that was used to extend the regions from the dark matter to the gas. This
works as follows:
\begin{enumerate}
	\item For every dark matter particle not in a Lagrangian region 
	      in the initial conditions, find the nearest $n$ neighbours.
	\item Find among the neighbours the maximal Lagrangian region ID,
	      corresponding to the lowest mass $z=0$ halo.
	\item Assign the particle the same Lagrangian region ID.
\end{enumerate}
The choice to assign the particles to the lowest mass halo, rather than the
higher mass halo, was made to ensure that spurious transfer into the lower mass
halo was avoided wherever possible. This means that the expectation is that
with this metric the level of inter-Lagrangian transfer will increase with
respect to the fiducial Lagrangian region identification method. This results
with the particles given to the haloes of a higher mass showing negligible
deviation from the fiducial result (see Fig \ref{fig:smoothconv}).

Note how smoothing the Lagrangian regions does have the expected effect of
inducing more inter-Lagrangian transfer, and does increase the proportion of
baryons that are classified as retained as the Lagrangian regions are filled
out. Despite this, the overall trends with respect to halo mass remain, with
a significant (>20\%) contribution from gas from outside Lagrangian regions
in haloes.

\subsection{The sizes of Lagrangian regions}

In Fig. \ref{fig:bigtransferpic} we saw that there was a large amount of
gaseous matter inside haloes from outside any Lagrangian region. It may be
reasonable to assume that this gas corresponds to dark matter that is simply
sitting just outside of the halo edge, perhaps within the so-called
`splashback radius'. The estimates for this radius range between 0.8 and
1.5$R_{\rm vir}$ \citep{More2015, Diemer2017a}, and hence below we consider
the situation where we extend the region around the halo that contributes to
the Lagrangian region. This is done in the following way:
\begin{enumerate}
	\item For every halo, find its current virial radius $R_{\rm vir}$. This contains
	      all particles at redshift $z=0$ that we consider to be within the halo.
    \item Now consider a new radius, $R_{\rm vir} \leq R_{\rm LR} \leq 1.5
		R_{\rm vir}$, and find all dark matter particles within this region
		from the halo centre. These dark matter particles are now defined to
		lie within the Lagrangian region of that halo.
    \item ID match these particles in the initial conditions to define the new
        Lagrangian region, extending to the gas in the usual way.
\end{enumerate}
\begin{figure}
    \centering
    \includegraphics{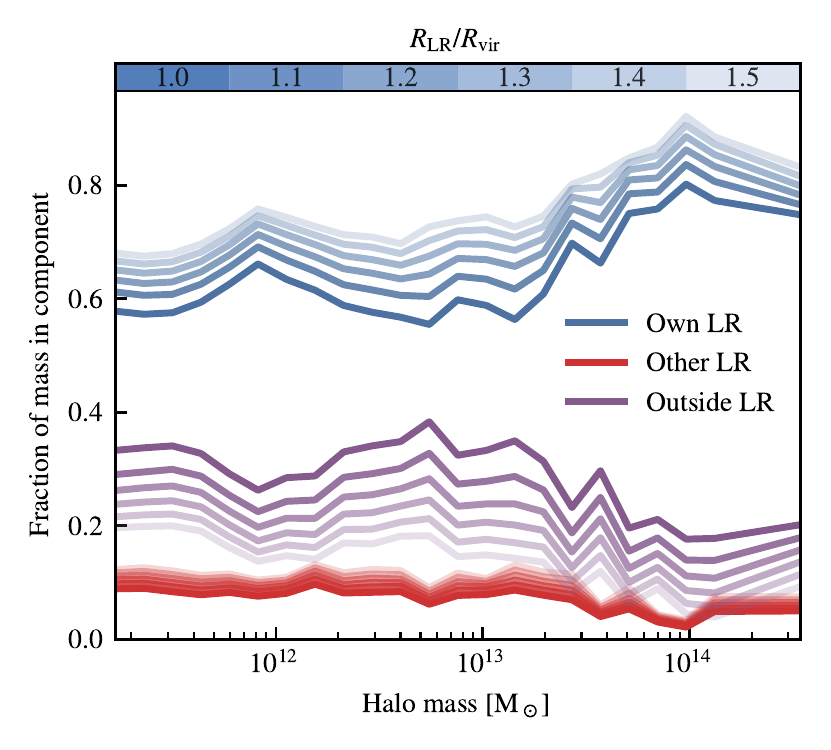}
    \vspace{-0.7cm}
    \caption{The same as Fig. \ref{fig:maintransferresult}, but now showing
    how the Lagrangian make-up of haloes is changed with an increasing radius
    for the definition of the Lagrangian region. Lighter colours correspond
    to larger radii, going in steps of $0.1R_{\rm vir}$ from 1.0 to 1.5.}
    \label{fig:radius_dependence}
\end{figure}
The effects of this process on the gas component of Fig.
\ref{fig:maintransferresult} (where it is most significant) are shown in Fig.
\ref{fig:radius_dependence}.Here we see that there is a significant change in
the fraction of mass in the halo at redshift $z=0$ from outside any
Lagrangian region, especially when going to $R_{\rm LR} = 1.5 R_{\rm vir}$.
This large change is expected, though, as we now have included a volume that
is three times larger than the initial halo in the Lagrangian region
classification; taking this extreme value for all haloes really is a
`worst-case' scenario. The inter-Lagrangian transfer remains at a similar
level despite the increase in radius. Note that there will be no extra mass
included in the haloes here, with particles simply changing their Lagrangian
allegiances.

We chose this specific process, increasing the radius of our Lagrangian
region rather than the whole halo, to prevent us from simply re-defining our
halo size and including more gas as well (as in this case, the transfer
across the halo boundary would simply be moved to a larger radius). 
\section{Discussion and Conclusions}
\label{sec:conclusions}

We have developed two novel metrics that describe the movement of baryons
throughout a cosmological simulation with respect to the dark matter, and
employed them to investigate the \simba{} simulations and sub-grid model.
The first of these metrics, the {\it spread metric}, shows that:
\begin{itemize}
    \item Dark matter can be spread up to $7.5\hmpc{}$ away from their inital
          mass distribution throughout the course of a cosmological simulation.
          This has been validated with two simulation codes, \gizmo{} and \swift{}.
    \item Gas can be spread to even larger distances, with the distance
          dependent on the physics included in the sub-grid model. For the
          \simba{} galaxy formation model with AGN jets, we find that gas can
          be spread to up to $12\hmpc{}$ throughout the course of the
          simulation in a box that is only $50\hmpc{}$ in size, with 40\%
          (10\%) of baryons having moved $> 1\hmpc{}$ ($3\hmpc{}$). This is
          despite this powerful form of feedback only directly interacting
          with 0.4\% of particles, and points towards significant quantities
          of gas being entrained by these jets. It remains to be seen if this
          will increase further with higher mass objects in larger boxes.
    \item Stars in the simulation show a very similar level of spread to the
          dark matter, suggesting that the gas particles that stars form out
          of remain tightly coupled to the dark matter. This implies that the
          spreading of stars by gravitational dynamics dominates over the
          spreading of their gas particle progenitors by feedback.
    \item Using the spread metric to select particles, we have shown that
          dark matter that is spread to large distances forms the diffuse
          structure within and around haloes, with lower spread dark matter
          forming substructure within haloes. When extending this to the gas,
          we find that the baryons that are spread the most are those that
          reside in the diffuse structure around haloes, with this structure
          being created by the energetic feedback present in the \simba{}
          model. We suggest that this spread metric may be a useful, highly
          computationally efficient, way of selecting particles that have been
          entrained by feedback processes that are not tagged during the
          injection of energy.
\end{itemize}
The second of these metrics, which considers the baryonic make-up of haloes
at $z=0$ split by the Lagrangian origin of the particles, shows that:
\begin{itemize}
    \item Approximately 40\% of the gas in an average $z=0$ halo did not originate
          in the Lagrangian region of that halo, with around 30\% originating
          outside any Lagrangian region, and 10\% originating in the Lagrangian
          region of another halo. This suggests that \emph{inter-Lagrangian
          transfer} is prevalent throughout the simulation, with haloes interchanging
          particles between $z=2$ and $z=0$ thanks to energetic feedback pathways.
    \item The majority of the stellar component of haloes (90\% above a halo mass
          $10^{12}\msolar{}$) originates from the Lagrangian region of the
          same halo, as expected given the similar large-scale spreads of the
          stellar and dark matter.
    \item Below a halo mass of $10^{13}\msolar{}$, haloes can only retain
          approximately 20-30\% of the baryons from their Lagrangian region,
          with the majority of these baryons being lost to the IGM. Above
          this mass, haloes become strong enough gravitational wells to
          retain the majority of their baryons (up to 60\%) by around
          $10^{14}\msolar{}$ halo mass, although this result is somewhat
          uncertain due to the lack of objects in this mass range in the
          $50\hmpc{}$ simulation box used here.
    \item Haloes with mass $M_H > 10^{13.5}\msolar{}$, despite having a baryon
          fraction comparable to the cosmic mean, still show significant
          levels of transfer from other haloes and from outside any
          Lagrangian region. This suggests a complex cycling of baryons with
          approximately 20\% of their baryonic mass being `swapped' with the
          IGM by $z=0$.
    \item Different Lagrangian components, as they make up the baryon
          fraction of haloes, are affected differently by feedback mechanisms
          at different halo masses. In the halo mass range
          $10^{12}$--$10^{13}\msolar{}$, the component of baryonic mass from
          outside of the Lagrangian region is halved, whereas the component
          from the haloes own Lagrangian region is only reduced within 20\%;
          this highlights the importance of preventive feedback for the
          baryon fraction of haloes.
\end{itemize}

Our results add a new perspective to the connection between baryon cycling
and galaxy evolution. Using large volume simulations including
momentum-driven winds, \citet{Oppenheimer2010} showed that most stars likely
form out of gas that has previously been ejected in winds, and more recent
zoom-in simulations agree with the prevalence of wind recycling
\citep{Christensen2016, AnglesAlcazar2017, Tollet2019}. Using the FIRE
simulations, \citet{AnglesAlcazar2017} further showed that the intergalactic
transfer of gas between galaxies via winds can provide up to a third of the
stellar mass of Milky Way-mass galaxies. Here we have introduced the concept
of inter-Lagrangian transfer, which represents the extreme case of transfer
of baryons between individual central haloes. For the \simba{} simulations,
we find that only a small fraction (<5\%) of the stellar mass of haloes can
be made up from inter-Lagrangian transfer gas, suggesting that most
intergalactic transfer originates from satellite galaxies and is thus
confined within Lagrangian regions. It is nonetheless quite significant that
gas exchanged between Lagrangian regions can fuel star formation in a
different halo at all. In addition, we do find a significant contribution
(<20\%) of inter-Lagrangian transfer to the gas content of haloes at z = 0.
Recently, \citet{Hafen2019} has highlighted the contribution of satellite
winds to the gas and metal content of the CGM in the FIRE simulations. Our
results suggest that the origin of the CGM of galaxies is linked to larger
scales than previously considered.

These results provide two possible main implications for current works. The
first is the implications for semi-analytic models of galaxy formation. These
models, by construction, tie the baryonic matter to dark matter haloes; they
contain no prescription for gas that explicitly originates from regions where
the dark matter does not end the simulation in a bound object. Also, whilst
there has been some effort by \citet{Henriques2015, White2015} and others to
include wind recycling into these models, there is currently no semi-analytic
model that includes any concept of baryon transfer between un-merged haloes
or baryonic accretion rates significantly different to that expected from the
dark matter component.

The second implication is for zoom-in simulation suites. These suites
typically construct their initial conditions by considering the cubic volume,
ellipsoid, or convex hull in the initial conditions containing the dark
matter particles that are located within a given distance (typically
$2-3R_{\rm vir}$) of the selected halo at $z=0$ \citep[see e.g.][]{Onorbe2014}.
However, our results highlight that the shapes of the
causally connected regions in gas and dark matter may be significantly
different. For example, the Latte \citep{Wetzel2016} suite uses an exclusion
region for high resolution particles of around $1.5 \hmpc{}$ while we find
that 10\% of cosmological baryons can move >3 Mpc away relative to the
original neighbouring dark matter distribution. While zoom-in simulations are
constructed to avoid contamination of low-resolution particles into the
high-resolution region, our results suggest that they may miss a flux of
external baryons into the high resolution region. In practice, contamination
from external sources will be somewhat mitigated by the usual choice of
isolated haloes, but future work should consider these effects for zoom-in
suites that have a full hydrodynamical simulation for their parent.

The results presented here are based on the \simba{} model, which is in good
agreement with a wide range of galaxy \citep{Dave2019} and black hole
\citep{Thomas2019} observables, but are clearly dependent on the feedback
implementation. Other galaxy formation models may yield different results,
especially those with drastically different implementations for AGN feedback,
such as the purely thermal feedback in the EAGLE model
\citep{Schaye2015}. The spread metric represents a unique tool to
characterize the global effects of feedback and will enable novel comparisons
between existing cosmological simulations. Future work should also address
the connection between baryon spreading and galaxy/CGM observables, as well
as investigate baryonic effects on cosmological observables
\citep{Schneider2015, Chisari2018} in the context of the spread metric.

\section{Acknowledgements}
\label{sec:acknowledgements}

The authors would like to thank James Willis for his help with the
\velociraptor{} halo finder, and Aaron Ludlow, Cedric Lacey, Richard Bower,
Shy Genel, Greg Bryan, and Rob Crain for helpful discussions that contributed
significantly to this work.

This work was initiated as a project for the Kavli Summer Program in
Astrophysics held at the Center for Computational Astrophysics of the
Flatiron Institute in 2018. The program was co-funded by the Kavli Foundation
and the Simons Foundation. We thank them for their generous support.

This work was supported by collaborative visits funded by the Cosmology and
Astroparticle Student and Postdoc Exchange Network (CASPEN).

JB is supported by STFC studentship ST/R504725/1. DAA is supported by the
Flatiron Institute, which is supported by the Simons Foundation. This work
used the ARCHER UK National Supercomputing Service.

This work used the DiRAC@Durham facility managed by the Institute for
Computational Cosmology on behalf of the STFC DiRAC HPC Facility
(www.dirac.ac.uk). The equipment was funded by BEIS capital funding via STFC
capital grants ST/K00042X/1, ST/P002293/1, ST/R002371/1 and ST/S002502/1,
Durham University and STFC operations grant ST/R000832/1. DiRAC is part of
the National e-Infrastructure. We would like to extend our thanks specifically
to Alastair Basden and his team for managing the DiRAC Memory Intensive service. 

\subsection{Software Citations}

This paper made use of the following software packages:
\begin{itemize}
    \item GIZMO \citep{Hopkins2017}
        \begin{itemize}
            \item Gadget \citep{Springel2005b}
        \end{itemize}
    \item {\sc Swift} \citep{Schaller2016}
    \item {\tt python} \citep{Rossum1995}, with the following libraries
    \begin{itemize}
    	\item {\tt numpy} \citep{Numpy2006}
    	\item {\tt scipy} \citep{Scipy2001}
    	\item {\tt py-sphviewer} \citep{Benitez-Llambay2015}
    	\item {\tt caesar} \citep{Thompson2018}
    	\item {\tt yt} \citep{Turk2011}
    \end{itemize}
    \item \velociraptor{} \citep{Elahi2019}
    \item The Amiga Halo Finder (AHF) \citep{Gill2004, Knollmann2009}
\end{itemize}

\bibliographystyle{mnras}
\bibliography{bibliography}

\end{document}